\documentclass[a4paper,11pt]{article}

\usepackage{amsmath,amssymb}
\usepackage{graphicx}
\usepackage{hyperref}
\usepackage{geometry}
\usepackage[noadjust]{cite}
\newcommand{\gev}{\,{\rm GeV}}
\newcommand{\tev}{\,{\rm TeV}}
\newcommand{\vevx}[1]{\left<{#1}\right>}

\newcommand{\mc}{\mathcal}
\newcommand{\tr}{\mathop{\mathrm{tr}}\nolimits}
\newcommand{\Tr}{\mathop{\mathrm{Tr}}\nolimits}
\newcommand{\Pf}{\mathop{\mathrm{Pf}}\nolimits}
\newcommand{\diag}{\rm{diag}}

\title{Safety of dynamical SUSY breaking and gaugino condensation in models of spontaneous CP violation}
\author{Norimi Yokozaki\thanks{n.yokozaki@gmail.com}\\[1mm]
  {\small Zhejiang Institute of Modern Physics and Department of Physics,}\\
  {\small Zhejiang University, Hangzhou, Zhejiang 310027, China}}
\date{}

\begin{document}
\maketitle

\begin{abstract}
\noindent
Solving the strong CP problem with spontaneous CP violation (SCPV) in SUSY requires strong
suppression of the phases transmitted by the constant term of the superpotential and the
SUSY-breaking $F$-term. When both are generated dynamically, by gaugino condensation and by dynamical SUSY
breaking (DSB), their phases need not vanish even with real couplings and CP-invariant
hidden $\theta$ angles. Both condensates of pure $SU(2)$ super Yang--Mills are real
at $\theta=0$ and purely imaginary at $\theta=\pi$. In massless SQCD with an
Affleck--Dine--Seiberg superpotential, the condensate phase depends on meson stabilization.
The phase of the vacuum superpotential enters the gluino mass through anomaly mediation.
With an explicit $\mu$ term in the superpotential, this phase can also enter the quark mass
determinant through the Higgs bilinear. The SUSY-breaking $F$-term is the main source of the MSSM soft masses, including the gluino mass, so its phase can reach $\bar\theta$ as well. The 3-2 model can mediate supersymmetry breaking through an $F$-term that is real along the R-axion orbit. The ISS model is safe in that sector as long as the quark-mass splittings exceed the messenger-loop corrections; when they fall below, a complex vacuum orientation can generate an $\mathcal{O}(1)$ phase. The IYIT model can generate a purely imaginary $F$-term: for one class of real Yukawa couplings, the meson VEV is purely imaginary and CP is spontaneously broken. In all three, a complex constant term can shift the VEV of the field coupled to the messengers and thereby induce a phase in the messenger mass, providing a second route to the strong CP angle.
\end{abstract}

\section{Introduction}
\label{sec:intro}

In addition to the usual renormalizable terms, the QCD Lagrangian admits a topological
$\theta$-term $(\theta g^2/32\pi^2)\,G_{\mu\nu}^a\tilde G^{a\mu\nu}$, which violates CP. The
physical phase in the Standard Model is the combination of $\theta$ and the phase of the
quark mass matrix determinant,
\begin{equation}
  \bar\theta\equiv\theta+\arg\det m_q .
  \label{eq:thetabar-def}
\end{equation}
A nonzero $\bar\theta$ generates a neutron electric dipole moment (EDM),
$d_n\simeq(1.5\text{--}2.4)\times10^{-16}\,\bar\theta\ e\cdot{\rm cm}$ from QCD sum rules
\cite{Pospelov:1999mv} and lattice QCD \cite{Dragos:2019oxn,Liang:2023jfj}, consistent with
the original chiral Lagrangian estimate \cite{Crewther:1979pi}.
The experimental bound on the neutron EDM, $|d_n|<1.8\times10^{-26}\,e\cdot{\rm cm}$
\cite{Abel:2020pzs}, therefore requires $\bar\theta\lesssim10^{-10}$. $\theta$ enters physical
quantities only through nonperturbative dynamics (instantons) \cite{tHooft:1976rip}. Since
$U(1)_A$ is anomalous, $\theta$ and $\arg\det m_q$ are separately unphysical, and
$\bar\theta$ cannot be removed by any field redefinition. It is a dimensionless angle of
period $2\pi$ protected by no symmetry of the Standard Model, so a generic value is $O(1)$,
and nothing forces it below $10^{-10}$ --- this is the strong CP problem.

Two well-known solutions exist. One is the Peccei-Quinn mechanism
\cite{Peccei:1977hh,Peccei:1977ur}, which promotes $\theta$ to the VEV of a dynamical
pseudoscalar field (the axion \cite{Weinberg:1977ma,Wilczek:1977pj}) so that the QCD
potential itself is minimized at $\bar\theta=0$ automatically. The other, the subject of
this paper, is spontaneous CP violation (SCPV): no axion is introduced, and CP is
instead taken to be an exact symmetry of the UV Lagrangian, broken only spontaneously by a
VEV that also generates the CP phase of the CKM matrix, thereby solving the strong CP
problem.

The general motivations for introducing SUSY --- resolving the gauge hierarchy problem
through the absence of quadratic divergences, consistency with the GUT scale suggested by
gauge coupling unification, and so on --- exist independently of the strong CP problem.
SUSY can play an analogous role in the context of the strong CP problem as well. Axion
models add a new scalar field responsible for the breaking of Peccei-Quinn symmetry and
its scale $f_a$; SCPV/Nelson-Barr constructions instead add a CP-violating gauge-singlet
scalar and its VEV scale. SUSY is a promising framework for keeping these scalar potentials
radiatively stable and for naturally generating the required scale hierarchies
\cite{Hiller:2001qg,Hiller:2002um,Nakagawa:2024ddd,Liu:2025ycm}.

This paper focuses on the SUSY Nelson-Barr construction
\cite{Nelson:1983zb,Barr:1984qx,Dine:1993qm,Evans:2020vil,Hisano:2025tud}. The Standard Model quarks are coupled to heavy
vector-like quarks, and a complex phase originating from the VEV of a gauge-singlet scalar
is mixed into an otherwise real UV mass parameter. The determinant structure of the full
quark mass matrix removes the heavy-light mixing phase from its tree-level determinant
\cite{Dine:1993qm,Dine:2015jga}. The phases of the Higgs vacuum expectation values
enter the determinant separately.

Throughout this paper we assume exact CP in the UV and take
\begin{equation}
  \theta_{{\rm bare},{\rm QCD}}=0,\qquad
  \theta_{{\rm bare},i}\in\{0,\pi\}\pmod{2\pi}\quad (i\text{ hidden}),
  \qquad \text{all Lagrangian couplings real}.
  \label{eq:setup}
\end{equation}
The index $i$ labels the hidden gauge groups. CP sends each bare angle to its negative
modulo $2\pi$, and both hidden-angle choices are CP invariant. CP violation thus arises
spontaneously.
We also assume that the SCPV
sector is sequestered from the SUSY-breaking and messenger sectors. The question is not one
of enumerating explicit phases in the Lagrangian (explicit CP violation), but whether
strongly-coupled nonperturbative dynamics generates a dangerous phase.

The Nelson-Barr protection concerns the quark mass matrix; it does not by itself guarantee
that the SUSY-breaking sector introduces no phase. If the gluino mass $M_3$ becomes
complex, its phase flows directly into $\bar\theta$ through the chiral anomaly. For a
bare Higgs mass parameter, the superpotential phase also enters $\bar\theta$ through
the Higgs bilinear and the resulting vacuum phase. In this paper we assume that the
constant term required for the vanishing cosmological constant is supplied by
gaugino condensation, and that the SUSY-breaking $F$-term is supplied by DSB --- so that
neither scale is put in by hand, but is instead explained dynamically through dimensional
transmutation. Anomaly mediation transmits the phase of the vacuum superpotential to
$M_3$. Gauge mediation transmits the phase of the $F$-term in the selected
SUSY-breaking vacuum. Depending on their relative size the phase from the
former can be diluted, whereas the phase from the latter is not diluted unless
the phases of $M_3$ and $b$ are aligned. In other words,
choosing gauge mediation does not by itself guarantee $\bar\theta\simeq0$.
The quantitative treatment of these two channels and their dilution is given in
Sec.~\ref{sec:M3channels}. The constant term can also shift the vacuum expectation value
of the field that couples the DSB sector to the messengers. A complex constant term can
thus transmit a phase through the messenger mass and, for some portals, through the
$F$-term itself. In IYIT and ISS the induced messenger mass phase persists at fixed gluino
mass as the messenger scale is lowered (Secs.~\ref{sec:M3channels}, \ref{sec:IYIT},
\ref{sec:ISS}, \ref{sec:32model} and \ref{sec:conclusion}).

A complete model of this type, combining the Nelson-Barr mechanism with gauge mediation and
with leptogenesis, has been constructed in \cite{Evans:2020vil}. There the constant term
required for the (nearly) vanishing cosmological constant comes from a tree-level superpotential whose parameters
are real by CP, and the SUSY-breaking sector is specified at tree level as well; the gluino
mass then comes out real at each of the three degenerate minima of the resulting potential.
In the language used here that construction lies in the safe class throughout, and the
scale of the constant term and the scale of SUSY breaking are both inputs. The present
paper asks what survives when both are instead generated by dimensional transmutation.

Moreover, imposing Eq.~(\ref{eq:setup}) does not by itself guarantee phase suppression. The
DSB sector is, by its nature, a strongly-coupled confining gauge theory, and a confining
$\mc{N}=1$ gauge theory generically has a complex vacuum degeneracy tied to an
anomalous $U(1)_R$ symmetry. That is, the very motivation
for using strong-coupling dynamics --- explaining scales through dimensional
transmutation --- is the same source that generates the danger of a CP phase. The
question to ask is therefore not whether to use strong coupling, but rather which
kind of strong-coupling dynamics can supply a scale without introducing a phase.
We examine the branch structure of the holomorphic solution, the phase of the vacuum
selected by the effective potential, and any phases introduced by holomorphic matching.
In pure SYM a root of the holomorphic scale yields isolated
condensate vacua, whose phases depend on the bare angle. In massless SQCD the ADS
superpotential also depends on the meson, and the phase selected in the vacuum depends on
the interactions that stabilize it. What safety in this sense delivers, namely reality up to an overall
sign, is discussed in Sec.~\ref{sec:M3channels}, and the criterion itself is worked out in
Secs.~\ref{sec:VY}--\ref{sec:ADS}.

This paper is organized as follows. Section~\ref{sec:M3channels} sets up the two channels
through which a phase reaches the gluino mass and the resulting bounds.
Section~\ref{sec:gaugino} derives the branch phase of gaugino condensation, and
Sec.~\ref{sec:Nc-constraint} shows that the quantum moduli constraint of $N_f=N_c$ SQCD
carries none. Sections~\ref{sec:IYIT}, \ref{sec:ISS} and \ref{sec:32model} examine the IYIT,
ISS and 3-2 models, and Sec.~\ref{sec:conclusion} summarizes the results.

\section{Phase channels to the gluino mass from gaugino condensation and DSB, and the safety criterion}
\label{sec:M3channels}

In the Nelson--Barr basis the bare QCD angle vanishes by Eq.~(\ref{eq:setup}).
We choose the real mass signs so that $\bar\theta=0$ in the real reference vacuum.
The triangular Nelson--Barr mass matrix removes the heavy-light mixing phase from
its determinant. For three generations the light-quark determinant contains
$(v_uv_d)^3$, where $v_u$ and $v_d$ are the neutral Higgs vacuum expectation values.
Including the gluino anomaly and the heavy colored thresholds gives
\begin{equation}
  \bar\theta=3\arg(v_uv_d)+3\arg M_3+\Delta\theta_{\rm other}
  \pmod{2\pi}.
  \label{eq:thetabar-M3}
\end{equation}
Here $M_3$ is the total gluino mass. The term $\Delta\theta_{\rm other}$ includes
the phase of the light-quark mass determinant after factoring out the Higgs VEVs,
as well as the heavy colored threshold contributions. The phases in
Eq.~(\ref{eq:thetabar-M3}) are measured relative to the reference signs,
with each heavy threshold included once. The experimental requirement
$|\bar\theta|\lesssim10^{-10}$ follows from the neutron EDM.

In this paper we assume that the constant term is supplied by gaugino condensation
and the SUSY-breaking $F$-term by DSB. We write
\begin{equation}
  \vevx{W}=\mc{C}+\vevx{W_{\rm other}},\qquad
  \mc{C}\equiv\vevx{W_{\rm cond}}\propto\vevx{\lambda\lambda},
  \label{eq:C-source}
\end{equation}
where $\mc{C}$ is the contribution from gaugino condensation. We assume that
$|\vevx{W_{\rm other}}|\ll|\mc{C}|$ and that $\vevx{W_{\rm other}}$ is real in
the common CP basis to the accuracy required by the strong CP bounds.
We therefore use $\vevx{W}\simeq\mc{C}$ in the benchmark estimates below.\footnote{%
Both scales can in principle be produced without strong-coupling dynamics. If $\mc{C}$
is taken to be a bare real constant placed in the superpotential by hand, and SUSY is
broken by a tree-level O'Raifeartaigh model with parameters likewise put in by hand, then
exact CP fixes everything to be real and none of the following dangers arise. In that case,
however, both the scale of $\mc{C}$ and the scale of SUSY breaking are inputs.} Throughout,
$M_P$ denotes the reduced Planck mass, $M_P\equiv(8\pi G_N)^{-1/2}\simeq2.4\times10^{18}\gev$.
The gravitino mass term of the supergravity Lagrangian reads
$m_{3/2}=e^{\vevx{K}/2M_P^2}\,\overline{\vevx{W}}/M_P^2$. In the
condensate-dominated approximation with $|\vevx{K}|\ll M_P^2$, this gives
$|\mc{C}|\simeq |m_{3/2}|M_P^2$, far below $M_P^3$. The reason is
common to both scales, namely that of explaining the scale dynamically. The scale
of SUSY breaking must be widely separated from $M_P$. The dimensional transmutation
of a strongly-coupled gauge theory, $\Lambda\sim M_P\exp[-8\pi^2/(b\,g^2)]$, explains
both hierarchies from an $O(1)$ gauge coupling. This choice is distinctly natural in that it
requires no small parameter put in by hand.

Under this assumption there are two channels through which a phase can reach $M_3$, and
they belong to separate sectors of different origin. The gaugino condensation
supplying $\mc{C}$ takes place in an $SU(N)$ of its own, distinct from the gauge group of
the DSB model that supplies the SUSY-breaking $F$-term, and the two sectors share no
fields. The first channel (A) is anomaly mediation
\cite{Giudice:1998xp,Randall:1998uk}, which generates
\begin{equation}
\begin{split}
  M_3^{\rm AMSB}=-\frac{g_3^2}{16\pi^2}\biggl[&(3T_G-T_R)\,m_{3/2}
  +(T_G-T_R)\,K_iF^i\\
  &+\frac{2T_R}{d_R}\bigl(\log\det K|_{R''}\bigr)_{,i}F^i\biggr]
  \qquad(\text{in units of }M_P=1)
\end{split}
  \label{eq:BMP}
\end{equation}
\cite{Bagger:1999rd}. Here $T_G$ is the Dynkin index of the adjoint and $T_R$ that of the
matter representation of dimension $d_R$, summed over all representations; for $SU(3)_c$
with the MSSM matter content, $T_G=3$, $T_R=6$ and $d_R=3$. $F^i$ is the vacuum expectation
value of the auxiliary field in the Einstein frame. The first term is pure anomaly
mediation, fixed by the conformal anomaly and depending on the SUSY-breaking sector only
through $m_{3/2}\propto\overline{\vevx{W}}$. The
second and third involve the K\"ahler potential and the auxiliary fields of that sector and
carry $\arg F^i$ instead \cite{Bagger:1999rd}; they become comparable to the first when
$\vevx{K_i}\sim1$ in Planck units, i.e.\ when the SUSY-breaking field has a VEV of order the
Planck scale. For the natural gauge-mediation case $\vevx{K_i}\ll1$, Eq.~(\ref{eq:BMP})
reduces to
\begin{equation}
  M_3^{\rm AMSB}=-\frac{g_3^2}{16\pi^2}\,b_0\,m_{3/2},\qquad b_0=3T_G-T_R=3
  \label{eq:AMSB}
\end{equation}
whose phase is $\pi-\arg\vevx{W}$, the $\pi$ coming from the overall sign.
This contribution to the strong CP phase was identified in Ref.~\cite{Dine:2015jga}.

The contribution $\vevx{W_{\rm SCPV}}$ is part of $\vevx{W_{\rm other}}$
even when the SCPV sector is sequestered from the DSB and messenger sectors.
We assume that its imaginary part in the coupled vacuum vanishes or is
sufficiently small to satisfy the strong CP bounds below. In the isolated SCPV sector,
an unbroken $R$ symmetry under which the superpotential transforms
nontrivially can enforce $\vevx{W_{\rm SCPV}}=0$ even when $R$-neutral fields
acquire complex VEVs \cite[Secs.~4 and 5.1.1]{Dine:2015jga}.

\paragraph{The Higgs contribution}
Use $H_u H_d=H_u^+H_d^- -H_u^0H_d^0$ and
$V\supset -(b H_u^0H_d^0+{\rm h.c.})$, with $b=B\mu$. At the neutral
tree-level minimum $\arg(v_uv_d)=-\arg b$, so the Higgs and gluino terms
in Eq.~(\ref{eq:thetabar-M3}) combine to $3\arg(M_3/b)$.
For a bare real $\mu$,
canonical rescaling leaves the superpotential term
$\int d^2\theta\,(1+\theta^2 F_{\rm comp})\mu H_uH_d$,
so $\delta b=-\mu F_{\rm comp}$ in this convention
\cite{Giudice:1998xp}. In the sequestered limit with negligible
$K_iF^i/M_P^2$, $F_{\rm comp}\simeq m_{3/2}$. In the low-scale
gauge-mediation regime considered here, this tree-level contribution is
a small perturbation of the weak-scale bilinear $b_{\rm r}$ in the real
reference vacuum. For example, in minimal gauge mediation the weak-scale
$B$ term is generated by renormalization-group evolution driven by the
gaugino masses, even with a vanishing gauge-mediated boundary value at
the messenger scale \cite[Sec.~IV]{Choi:2020wdq}. Defining
$B_{\rm r}=b_{\rm r}/\mu$, the fractional compensator correction is
$|\delta b/b_{\rm r}|=|F_{\rm comp}/B_{\rm r}|\ll1$ in this regime.
We use $|B_{\rm r}|=100$--$1000\,\mathrm{GeV}$ as a weak-scale benchmark.

Expanding about the reference vacuum with $b=b_{\rm r}+\delta b$ and
$M_3=M_{3,\rm r}+\delta M_3$ gives
\begin{equation}
 \delta\bar\theta=3\,\mathrm{Im}\!\left(
 \frac{\delta M_3}{M_{3,\rm r}}-\frac{\delta b}{b_{\rm r}}\right)
 +\delta(\Delta\theta_{\rm other}).
 \label{eq:Higgs-phase}
\end{equation}
The direct compensator contribution to the Higgs term is
$\delta\bar\theta_H\simeq3\,\mathrm{Im}\,F_{\rm comp}/B_{\rm r}$.
It can exceed the gluino term because the latter carries the anomaly-mediation
loop factor. In the absence of cancellations, $|\bar\theta|\lesssim10^{-10}$
then requires $|\mathrm{Im}\,F_{\rm comp}|\lesssim
33\,\mathrm{eV}\,(|B_{\rm r}|/1\,\mathrm{TeV})$, or about
$3$--$33\,\mathrm{eV}$ for the benchmark range above.

A common gauge-mediated phase of $M_3$ and $b$ cancels in $3\arg(M_3/b)$ at leading
order. For an order-one common phase and a bare $\mu$ term, even a real compensator
contribution to $b$ can shift the relative phase enough to exceed the strong CP bound.

The second channel (B) is the SUSY-breaking $F$-term. Throughout we write $Z_{\rm eff}$ for
the field that breaks SUSY and $F_{Z_{\rm eff}}$ for its $F$-term: $X$ and $F_X$ in
Sec.~\ref{sec:IYIT}, $\Phi_0$ and $F_\Phi$ in Sec.~\ref{sec:ISS}, $X_1$ and $F^{X_1}$ in
Sec.~\ref{sec:32model}. Through the messengers $F_{Z_{\rm eff}}$ generates
\begin{equation}
  M_3^{\rm GMSB}=\frac{g_3^2}{16\pi^2}\,N_{\rm mess}\,\Lambda_{\rm mess}
  =\frac{g_3^2}{16\pi^2}\,N_{\rm mess}\,\frac{\eta_{\rm eff}\,F_{Z_{\rm eff}}}{M_{\rm mess}} .
  \label{eq:GMSB}
\end{equation}
Here $N_{\rm mess}\equiv\sum_r 2T_3(r)$ is the messenger index, the sum running over the
messenger pairs in representations $r$ of $SU(3)_c$ with a common mass, normalized so that a
pair $\mathbf3\oplus\bar{\mathbf3}$ (e.g.\ from $\mathbf5\oplus\bar{\mathbf5}$) counts as 1, and $\eta_{\rm eff}$
is the model-dependent constant relating the messenger $B$-term to $F_{Z_{\rm eff}}$, i.e.\ it is
determined by how $F_{Z_{\rm eff}}$ is transmitted to the messenger sector. It is the
coefficient of the operator that gives the messengers their mass, so it is also the
coupling through which $\vevx{Z_{\rm eff}}$ reaches them. It is smaller the higher the
dimension of that operator: with the
renormalizable portal $W\supset\kappa Xf\bar f$ of Sec.~\ref{sec:IYIT} one has
$\eta_{\rm eff}=\kappa=O(1)$, with the composite portal of Sec.~\ref{sec:ISS} one power of
the ratio of the dynamical scale to the cutoff of that operator, and with that of
Sec.~\ref{sec:32model} two powers. If $M_{\rm mess}$ is real, then $\arg M_3=\arg(\eta_{\rm eff}\,F_{Z_{\rm eff}})$, and the phase of $F_{Z_{\rm eff}}$ becomes the
phase of $M_3$ directly. Unlike (A), however, this phase is fixed only once
the minimization of $V_{\rm eff}$ determines where the SUSY-breaking vacuum sits.

The (nearly) vanishing cosmological constant also ties the two channels together. Vanishing vacuum
energy, $\vevx{|F_{Z_{\rm eff}}|^2-3|W|^2/M_P^2}=0$, requires
\begin{equation}
  |F_{Z_{\rm eff}}|^2\simeq\frac{3|\mc{C}|^2}{M_P^2}=3\,|m_{3/2}|^2M_P^2 ,
  \label{eq:CC-condition}
\end{equation}
so the scale of the DSB sector is fixed
once the scale of the sector supplying $\mc{C}$ is, and the two strongly-coupled sectors
cannot be chosen independently. In terms of each model's own scales, up to couplings,
\begin{equation}
  |F_{Z_{\rm eff}}|\sim\Lambda^2\ \ \text{(IYIT)},\qquad
  |F_{Z_{\rm eff}}|\sim m_0\Lambda\ \ \text{(ISS)},\qquad
  |F_{Z_{\rm eff}}|\sim\Lambda_3^2\ \ \text{(3-2)} ,
  \label{eq:FZ-models}
\end{equation}
so that in IYIT and in the 3-2 model the dynamical scale itself is fixed, while in ISS the
relation constrains the product of the quark mass and the dynamical scale, leaving $\Lambda$
free.

For a dominant positive real gauge-mediated gluino mass, its phase shift from
anomaly mediation is suppressed by the anomaly-mediated fraction. Using
$\vevx{W}\simeq\mc{C}$ gives
\begin{equation}
  \arg M_3^{\rm tot}\simeq r_{\rm AMSB}\,\sin\bigl(\arg\mc{C}\bigr),
  \qquad r_{\rm AMSB}\equiv\frac{|M_3^{\rm AMSB}|}{|M_3^{\rm tot}|},
  \label{eq:dilution}
\end{equation}
and its contribution to Eq.~(\ref{eq:thetabar-M3}) is constrained by
\begin{equation}
  |\delta\bar\theta_{\tilde g}|
  \simeq3\,r_{\rm AMSB}\,|\sin(\arg\mc{C})|\lesssim10^{-10}
  \label{eq:C-quantitative}
\end{equation}
in the absence of cancellations.
Since Eqs.~(\ref{eq:AMSB}) and (\ref{eq:GMSB}) share the same loop factor $g_3^2/16\pi^2$,
their relative size is set by the scales multiplying it, namely the ratio of $|m_{3/2}|$ to
$|\Lambda_{\rm mess}|$, with $\Lambda_{\rm mess}=\eta_{\rm eff} F_{Z_{\rm eff}}/M_{\rm mess}$:
\begin{equation}
  r_{\rm AMSB}\simeq\frac{3\,|m_{3/2}|}{N_{\rm mess}|\Lambda_{\rm mess}|}
  =\frac{3\,|m_{3/2}|M_{\rm mess}}{N_{\rm mess}\,|\eta_{\rm eff}F_{Z_{\rm eff}}|}
  =\frac{\sqrt3}{N_{\rm mess}\,|\eta_{\rm eff}|}\,\frac{M_{\rm mess}}{M_P}
  \label{eq:rAMSB-gmsb}
\end{equation}
(from $\Lambda_{\rm mess}=\eta_{\rm eff}F_{Z_{\rm eff}}/M_{\rm mess}$, then
$|F_{Z_{\rm eff}}|=\sqrt3\,|m_{3/2}|M_P$). The $F$-term drops out, leaving
$M_{\rm mess}/M_P$ and $|\eta_{\rm eff}|$.
Compared with gravity mediation or pure AMSB, for which $r_{\rm AMSB}=O(1)$, gauge
mediation suppresses the phase from the constant term in the gluino mass. The phase
of $F_{Z_{\rm eff}}$ enters the dominant gauge-mediated mass directly through
Eq.~(\ref{eq:GMSB}). The Higgs contribution is governed by Eq.~(\ref{eq:Higgs-phase}).

For real $\mc{C}$, the assumed reality of $\vevx{W_{\rm other}}$ makes the
anomaly-mediated contribution real to the required accuracy. A generic phase
of $\mc{C}$, such as $2\pi/3$, can require substantial dilution to solve the
strong CP problem\footnote{%
When the requirement is not satisfied, this means only that ``solving the strong CP
problem via SCPV without an axion does not work'' --- it does not mean the model is
excluded. All claims in this paper are conditional in this sense.}. We now examine what is
phenomenologically required in that case.

With $|M_3^{\rm AMSB}|=3(g_3^2/16\pi^2)|m_{3/2}|$, the gluino benchmark in
Eq.~(\ref{eq:C-quantitative}) becomes
\begin{equation}
  |\delta\bar\theta_{\tilde g}|\simeq0.063\,\frac{|m_{3/2}|}{|M_3|}\,
  |\sin(\arg\mc{C})|\lesssim10^{-10}
  \label{eq:m32-bound}
\end{equation}
(with $g_3^2({\rm TeV})\simeq1.1$). The phase originating in the branch is
$O(1)$ in this benchmark. Normalizing to $|\sin(\arg\mc{C})|=\sqrt3/2$,
as for the branches $\arg\mc{C}=\pm2\pi/3$ of a pure $SU(3)$ condensate, gives
\begin{equation}
\begin{split}
  |m_{3/2}|&\lesssim(3.7\text{--}37)\ {\rm keV}\,
  \frac{\sqrt3/2}{|\sin(\arg\mc{C})|}\qquad(|M_3|=2\text{--}20\,\tev),\\[1mm]
  \sqrt{|F_{Z_{\rm eff}}|}&\lesssim(0.39\text{--}1.2)\times10^4\tev\,
  \left[\frac{\sqrt3/2}{|\sin(\arg\mc{C})|}\right]^{1/2} .
\end{split}
\label{eq:m32-F-bound}
\end{equation}
The range $|M_3|=2$--$20\,\tev$ spans the
LHC lower bound on the gluino mass at the low end. Except in special models with large
$A$-terms, the measured Higgs mass $m_h\simeq125\,\gev$ requires stops of order $10\,\tev$
or more in the MSSM; the gluino mass is then typically of the same order,
$|M_3|=O(10)\,\tev$. For a bare real $\mu$, an $O(1)$ compensator phase and
$|F_{\rm comp}|\simeq|m_{3/2}|$, the Higgs contribution in
Eq.~(\ref{eq:Higgs-phase}) gives an eV-scale upper bound on the
gravitino mass in the benchmark above, in the absence of cancellations.
Such a light gravitino is difficult to realize in perturbative gauge mediation
with a gluino mass of order $10\,\tev$.

A stable thermal gravitino contributes to the cosmological hot relic
density. In a standard thermal history one finds $|m_{3/2}|<2.3\,\mathrm{eV}$
\cite{Xu:2021rwg}, with related constraints in
\cite{Viel:2005qj,Osato:2016ixc,Peters:2023asu}. The Higgs mass and LHC
constraints give a lower scale of order $1\,\mathrm{eV}$ in the gauge-mediation
scenarios considered in \cite{Ibe:2016kyg,Lu:2017oee,Hook:2018sai}.
Constructions in this window have been explored \cite{Choi:2021vqf}.
Late entropy production can relax the bound on a thermal relic
\cite{Moroi:1993mb}. For example, we assume a small superpotential mixing
$\delta W=f_i(\vevx{W}/M_P^2)\,\mathbf{5}_M\overline{\mathbf{5}}_i$,
where $\mathbf{5}_M$ is a messenger multiplet and $\overline{\mathbf{5}}_i$ is an MSSM matter
multiplet \cite{Fujii:2002fv}. This mixing can arise from the K\"ahler potential
term $K\supset f_i\,\mathbf{5}_M\overline{\mathbf{5}}_i+\mathrm{h.c.}$, with an $R$-neutral bilinear.
This mixing allows the lightest messenger to decay.
If the messengers dominate the energy density and decay after gravitino decoupling, the
entropy released can reduce the thermal gravitino abundance enough to satisfy
cosmological bounds \cite[Appendix~B]{Hamaguchi:2014MDGM}.
The Higgs and gluino phase bounds apply independently of this dilution.

The messenger displacement gives a further contribution. In Secs.~\ref{sec:IYIT} and
\ref{sec:ISS}, where $Z_{\rm eff}$ is a pseudomodulus whose $F$-term is the order parameter,
the constant term shifts it to
\begin{equation}
  \vevx{Z_{\rm eff}}=c_Z\,\frac{16\pi^2\,\mc{C}}{\lambda_{\rm eff}^3M_P^2},
  \qquad c_Z\equiv\frac{2}{(\ln4-1)\,n_Z}
  \label{eq:Zeff-shift}
\end{equation}
(Eqs.~(\ref{eq:IYIT-Raxion-pot}) and (\ref{eq:ISS-C-tadpole}), with $\lambda_{\rm eff}$ the
canonically normalized Yukawa coupling of the DSB sector: $\lambda$ in Sec.~\ref{sec:IYIT},
up to the normalization of the composite meson, and $h$ in Sec.~\ref{sec:ISS}). The loop factor $16\pi^2$ is the enhancement of the shift over
$|m_{3/2}|\simeq|\mc{C}|/M_P^2$ caused by the one-loop origin of the pseudomodulus mass, and $n_Z$ is the
number of chiral multiplets that acquire the mass $\lambda_{\rm eff}Z_{\rm eff}$ and run in
that loop: $n_Z=5$ and $c_Z=1.04$ in Sec.~\ref{sec:IYIT}, and $n_Z=2\tilde N_c$ and
$c_Z=2.6/\tilde N_c$ in Sec.~\ref{sec:ISS}.
Then $\vevx{Z_{\rm eff}}$ enters the messenger mass alongside the bare mass,
Eqs.~(\ref{eq:IYIT-M3-mess}) and (\ref{eq:ISS-M3-mess}). In gauge mediation the magnitudes
are related: $|m_{3/2}|\simeq|F_{Z_{\rm eff}}|/(\sqrt3M_P)$ together with
$\Lambda_{\rm mess}=\eta_{\rm eff} F_{Z_{\rm eff}}/M_{\rm mess}$ gives
\begin{equation}
  \frac{|m_{3/2}|}{M_{\rm mess}}\simeq\frac{|\Lambda_{\rm mess}|}{\sqrt3\,|\eta_{\rm eff}|\,M_P} ,
  \label{eq:m32-over-Mmess}
\end{equation}
which does not depend on $M_{\rm mess}$. The coupling that transmits $F_{Z_{\rm eff}}$ to the
messengers also transmits $\vevx{Z_{\rm eff}}$, so $\eta_{\rm eff}$ cancels between the
two: it multiplies $\vevx{Z_{\rm eff}}$ in the messenger mass and divides it through
Eq.~(\ref{eq:m32-over-Mmess}). Let $\phi_{\rm mess}$ be the phase shift of the
common messenger mass $M_{\rm mess}+\eta_{\rm eff}\vevx{Z_{\rm eff}}$ relative
to the real reference vacuum. For real $F_{Z_{\rm eff}}$ and a dominant
gauge-mediated gluino mass, $\phi_{\rm mess}$ enters Eq.~(\ref{eq:thetabar-M3})
with weight $-3$ through the gluino and $+N_{\rm mess}$ through the colored
threshold. Including the response of the Higgs bilinear gives
\begin{equation}
  \delta\bar\theta_{\rm mess}\simeq
  (N_{\rm mess}-3)\phi_{\rm mess}-3(\delta\arg b)_{\rm mess}.
  \label{eq:shift-floor}
\end{equation}
Here $(\delta\arg b)_{\rm mess}$ is the change induced by the messenger phase.
For example, if all gauge-mediated gaugino masses acquire the common phase
shift $-\phi_{\rm mess}$ and the gaugino-driven RG contribution dominates $B$,
then $(\delta\arg b)_{\rm mess}\simeq-\phi_{\rm mess}$ and
$\delta\bar\theta_{\rm mess}\simeq N_{\rm mess}\phi_{\rm mess}$.
The direct compensator contribution to $b$ is included separately as in
Eq.~(\ref{eq:Higgs-phase}). To leading order in the messenger mass shift,
\begin{equation}
  |\phi_{\rm mess}|\simeq
  \frac{c_Z}{N_{\rm mess}}\,5.4\times10^{-12}\,
  \frac{|M_3|}{\tev}\,\frac{|\sin(\arg\mc{C})|}{\lambda_{\rm eff}^3}.
  \label{eq:messenger-phase-size}
\end{equation}
At fixed gluino mass, $M_{\rm mess}$ and the induced shift
$\eta_{\rm eff}\vevx{Z_{\rm eff}}$ scale together, so this leading
messenger phase is independent of the messenger scale.

We perform three independent checks (Sec.~\ref{sec:intro}). In (i) we examine
the branches of the holomorphic solution, their phases, and their connection
through continuous field directions. In (ii) we examine whether the effective
potential selects a real vacuum, including a metastable vacuum as in
Sec.~\ref{sec:ISS}. Loop corrections can select an orientation left degenerate
by the $F$-term potential. A pseudomodulus stabilized at the origin by the
Coleman--Weinberg potential can be displaced by $\mc{C}$, and explicit
$R$ breaking, such as an independent $\mc{C}$, can lift an R-axion direction.
In (iii) we examine the phases transmitted by holomorphic matching. The following
sections apply these checks to each model.

What the safe class delivers is reality: the strong-coupling dynamics generates a
phase in $\{0,\pi\}$, i.e.\ $\mc{C}$ and the $F$-term are real. Which of the two signs is
realized is fixed by model-dependent couplings outside the strong dynamics\footnote{%
The pure-SYM dynamics admits all the branches of Eq.~(\ref{eq:S-branch-SYM}).
For pure $SU(2)$ at $\theta=0$, the branch label fixes the sign of $\mc{C}$.
With $\Delta\theta_{\rm other}$ held at its reference value,
$\bar\theta=0$ requires the reference value of $\arg(M_3/b)$.
Reversing the relative sign of $M_3$ and $b$ shifts $\bar\theta$ by $\pi$ modulo $2\pi$.
The required relative sign is a separate assumption in the Nelson--Barr setup.}.

We assume that our observable Universe occupies a single vacuum. This can
occur if the relevant transition takes place sufficiently early during
inflation and the vacuum choice survives reheating. If distinct vacua remain
after the sectors are coupled and are populated in different regions after
inflation, domain walls can form. This applies to the condensate vacua of pure
SYM, including the two real vacua of pure $SU(2)$ at $\theta=0$, and to ADS
theories when explicit $R$ breaking produces multiple isolated vacua, as in the
full-rank quark-mass deformation. It also applies to the two real IYIT vacua
$\pm M$ and to the CP-conjugate pair in the purely imaginary IYIT class, as well
as to ISS when messenger effects select distinct CP-conjugate metastable vacua.

\section{Gaugino condensation and the phase of the constant term}
\label{sec:gaugino}

The scale $\Lambda$ that appears frequently below is the dynamical scale of a
strongly-coupled gauge theory, but the holomorphic physical quantity is not $\Lambda$
itself but rather $\Lambda^b$ ($b$ being the one-loop beta-function coefficient, $b=3N_c-N_f$
for SQCD and $b=3N_c$ for pure SYM):
\begin{equation}
  \Lambda^{b}\equiv\mu_0^{b}\exp\left[-\frac{8\pi^2}{g^2(\mu_0)}+i\theta\right]
  \qquad\Longrightarrow\qquad \arg\Lambda^{b}=\theta .
  \label{eq:Lambda-b-phase}
\end{equation}
The symbol $\Lambda$ itself is defined up to the choice of $b$-th root.
At the two CP-invariant angles in Eq.~(\ref{eq:setup}), $\Lambda^b$ equals
$+|\Lambda|^b$ at $\theta=0$ and $-|\Lambda|^b$ at $\theta=\pi$.
Taking an $N_c$-th root yields the phases of the pure-SYM condensates.

\subsection{Origin as a logarithmic superpotential}
\label{sec:VY}

The gaugino condensate phases arise from the low-energy effective
superpotential of pure $SU(N_c)$ super Yang-Mills theory (Veneziano-Yankielowicz, VY)
\cite{Veneziano:1982ah}, in the normalization of \cite{Intriligator:1995au},
\begin{equation}
  \begin{split}
  W_{VY}(S)&=\left[\left(-\frac{8\pi^2}{g^2(\mu_0)}+i\theta\right)S
  +N_cS\left(1-\ln\frac{S}{\mu_0^3}\right)\right]\\
  &=S\left[\ln\frac{\Lambda^{3N_c}}{S^{N_c}}+N_c\right],
  \qquad S\equiv-\frac{1}{32\pi^2}\tr W_\alpha W^\alpha .
  \end{split}
  \label{eq:WVY}
\end{equation}
The logarithmic term is required to reproduce the $U(1)_R$ anomaly. Since $S$ carries
$R$-charge 2, under a $U(1)_R$ rotation $S\to e^{2i\alpha}S$ one has
$\ln S\to\ln S+2i\alpha$, which produces, apart from the overall factor $e^{2i\alpha}$, the
extra term $-2iN_c\alpha S$. By the first line of Eq.~(\ref{eq:WVY}) this is equivalent to
$\theta\to\theta-2N_c\alpha$, which is exactly the anomalous shift induced by the gluino
rotation $\lambda\to e^{i\alpha}\lambda$. A polynomial superpotential cannot reproduce this anomalous
transformation property, so the log term is essential.

$\partial W_{VY}/\partial S=0$ gives the degree-$N_c$ polynomial equation
$S^{N_c}=\Lambda^{3N_c}$, with $N_c$ roots
\begin{equation}
  \vevx{S}_k=|\Lambda|^3\exp\!\left[\frac{i(\theta+2\pi k)}{N_c}\right],
  \qquad k=0,1,\dots,N_c-1
  \label{eq:S-branch-SYM}
\end{equation}
(that the right-hand side $\Lambda^{3N_c}$ carries no phase beyond that of
Eq.~(\ref{eq:Lambda-b-phase}) is confirmed by the direct computation of a single $\nu=1$
instanton in the massless theory, Sec.~\ref{sec:Nc-constraint}, Eq.~(\ref{eq:GN0}): the
invariant $\vevx{\lambda\lambda}^{N_c}=\gamma_{N_c}^0\Lambda^{3N_c}$ computed there has a
real coefficient $\gamma_{N_c}^0$). The multivaluedness is not
transmitted to the condensate directly --- what emerges is simply the $N_c$ roots of a
polynomial equation. It is in why the equation takes this form that the log term
matters, and this is the sense in which we call it ``logarithmic'' here. The same $W_{VY}$
reappears when $N_f=N_c$ SQCD is made massive and flowed to pure SYM in the deep IR
(Sec.~\ref{sec:mass-def}).

Substituting this extremum into Eq.~(\ref{eq:WVY}) gives the constant term. Since
$\vevx{S}_k^{N_c}=\Lambda^{3N_c}$ is independent of the branch $k$ ($e^{2\pi ik}=1$), the
argument of the logarithm ($=1$) is likewise branch-independent, and
\begin{equation}
  \mc{C}=W_{VY}(\vevx{S}_k)=N_c\vevx{S}_k .
  \label{eq:C-def}
\end{equation}
The phase of $\mc{C}$ is therefore precisely the phase of $\vevx{S}_k$.
Only the phase here matters below; $|\mc{C}|$ is tied to the SUSY-breaking scale by
Eq.~(\ref{eq:CC-condition}).
At $\theta=0$, the real branches are $k=0$ and, for even $N_c$, $k=N_c/2$.
At $\theta=\pi$, there is one real branch, $k=(N_c-1)/2$, for odd $N_c$,
and every branch is nonreal for even $N_c$. CP pairs $k$ with $-k$ at
$\theta=0$ and with $-k-1$ at $\theta=\pi$, modulo $N_c$.
In particular, pure $SU(2)$ has two real condensates at $\theta=0$ and
two purely imaginary condensates at $\theta=\pi$.

\subsection{Origin as a fractional-power (ADS-type) superpotential}
\label{sec:ADS}

Consider SQCD with $1\le N_f<N_c$ and vanishing tree-level masses for the
quark multiplets $Q^i,\tilde Q_i$. The classical flat directions allow the
meson VEV to vary continuously. Write $N'\equiv N_c-N_f$ and
$b\equiv 3N_c-N_f$.

This masslessness is not automatic. Since $Q^i,\tilde Q_i$ are vector-like under $SU(N_c)$,
gauge invariance alone does not forbid a mass term $m^i_jQ^i\tilde Q_j$. What protects
$m=0$ is instead the additional assumption that the anomaly-free chiral flavor
symmetry $SU(N_f)_L\times SU(N_f)_R\times U(1)_B\times U(1)_R$, which is enhanced at
$m=0$, is an exact symmetry of the UV Lagrangian (by a nonrenormalization theorem, $m=0$ is
then perturbatively stable if the symmetry is exact). The 3-2 model of
Sec.~\ref{sec:32model} has genuinely chiral matter content, so gauge invariance alone
forbids the mass term and this assumption is not needed.

At a generic point of the moduli space ($Q^i=\tilde Q_i=v_i\delta^i_a$ with
$a$ the color index, $M^i_j=v_i^2\delta^i_j$, $\det M\neq0$), $SU(N_c)$
is Higgsed to $SU(N')$ for $N'>1$. The unbroken gauge sector is pure
super Yang--Mills, with holomorphic matching
\begin{equation}
  \Lambda_{\rm low}^{3(N_c-N_f)}=\frac{\Lambda^{3N_c-N_f}}{\det M} .
  \label{eq:ADS-matching}
\end{equation}
Gaugino condensation in this sector gives $S_k=\Lambda_{\rm low}^3e^{2\pi ik/N'}$
and a contribution $N'S_k$ to the superpotential. For $N'=1$, the gauge group
is completely Higgsed and a single instanton generates the superpotential.
The result in both cases is
\begin{equation}
  W_{ADS}(M)=(N_c-N_f)\left(\frac{\Lambda^{3N_c-N_f}}{\det M}\right)^{\frac{1}{N_c-N_f}}
  e^{2\pi ik/(N_c-N_f)}
  \label{eq:WADS}
\end{equation}
\cite{Affleck:1983mk}. With $\phi=\arg\det M$ followed continuously along the
meson direction, the phase on sheet $k$ is
\begin{equation}
  \psi_k(\phi)\equiv\arg W_{ADS}=\frac{\theta-\phi+2\pi k}{N'} .
  \label{eq:ADS-sheet-phase}
\end{equation}
Winding $\phi$ by $2\pi$ at fixed $\theta$ takes sheet $k$ to sheet $k-1$,
so the sheets are connected along the meson field space. For $N'=1$ this
winding returns to the same sheet. In that case, at fixed real, nonzero
$\det M$, shifting $\theta$ from zero to $\pi$ reverses the sign of the
real $W_{ADS}$.

The anomaly-free $U(1)_R$ has $R(Q)=R(\tilde Q)=1-N_c/N_f$, so
$R(\det M)=-2N'$ and $W_{ADS}$ has charge 2. The supersymmetric potential
runs away toward $|\det M|\to\infty$. A positive $R$-invariant squark
soft mass can stabilize the radial direction while leaving the $R$ orbit
of the nonzero meson VEV flat.

An independent soft term with a coefficient $A$ of mass dimension one gives
\begin{equation}
  \Delta V=-2\operatorname{Re}(A^*W_{ADS})
  =-2|A W_{ADS}|\cos(\psi_k-\arg A).
  \label{eq:ADS-soft-phase}
\end{equation}
When this term dominates the angular potential, its minimum is at
$\psi_k=\arg A$ modulo $2\pi$, with curvature $2|A W_{ADS}|/(N')^2$ along
$\phi$ at fixed radial VEVs. Thus a real $A$ in the common CP basis fixes a
real condensate for any $N'$ at either $\theta=0$ or $\pi$. The ADS
superpotential's own contribution through the compensator is proportional to
$|W_{ADS}|^2$ and leaves the angular direction flat.

For a full-rank supersymmetric quark mass matrix $m_Q$, extremizing
$W=W_{ADS}+\Tr(m_QM)$ gives $m_QM=S\mathbf{1}$ with $S=W_{ADS}/N'$.
Taking the determinant gives $\det M=S^{N_f}/\det m_Q$, and the ADS
relation $S^{N'}=\Lambda^b/\det M$ yields $S^{N_c}=\Lambda^b\det m_Q$.
With the holomorphic matching $\Lambda_{\rm SYM}^{3N_c}=\Lambda^b\det m_Q$,
the effective superpotential in each of the $N_c$ isolated supersymmetric
vacua is
\begin{equation}
  W_{\rm eff}^{(k)}=N_c S_k
  =N_c\left(\Lambda^b\det m_Q\right)^{1/N_c}e^{2\pi ik/N_c},
  \qquad k=0,\ldots,N_c-1 .
  \label{eq:ADS-mass-deformation}
\end{equation}
In the convention $\det m_Q>0$, with $\theta$ defined in the same basis,
the matching gives $\arg\Lambda_{\rm SYM}^{3N_c}=\arg\Lambda^b=\theta$
modulo $2\pi$. The branch phases and their CP properties are then the same as in pure SYM,
Eq.~(\ref{eq:S-branch-SYM}).

\paragraph{Normalization of the coefficient}
The overall coefficient of Eq.~(\ref{eq:WADS}) depends on the definition of $\Lambda$ only
through real factors. With the holomorphic scale $\Lambda^{b}$ of
Eq.~(\ref{eq:Lambda-b-phase}) and its one-loop constant fixed in the
$\overline{\rm DR}$ scheme (equivalently Pauli-Villars or the $\zeta$-function scheme;
$\overline{\rm MS}$ differs by the real factor $e^{-I_2({\rm adj})/12}$, with
$I_2({\rm adj})=2N_c$ in the normalization in which the fundamental has index $1$), the threshold
matching in Eq.~(\ref{eq:ADS-matching}) is exactly 1 by Weinberg's background-field
one-loop threshold formula \cite{Weinberg:1980wa}, and the coefficient is exactly $N_c-N_f$
\cite{Finnell:1995dr}. Other conventions rescale both by real factors.

For $N'=1$ the coefficient has also been computed directly from the single instanton.
Cordes \cite{Cordes:1985um} carried this out for general $N_c$, and Finnell and Pouliot
\cite{Finnell:1995dr} corrected a factor of 4 in it, obtaining
\begin{equation}
  W_{SU(N_c)}=\frac{\Lambda^{2N_c+1}}{\det Q\tilde Q}
  \qquad(N_f=N_c-1),
  \label{eq:ADS-instanton}
\end{equation}
in agreement with Eq.~(\ref{eq:WADS}). The collective-coordinate integral at the completely
Higgsed, weakly-coupled point is real and positive and multiplies $\Lambda^{2N_c+1}$, which
is real at both CP-invariant values of $\theta$. This is the case used by the 3-2 model of Sec.~\ref{sec:32model}.

\section{Phase structure of the $N_f=N_c$ quantum moduli constraint}
\label{sec:Nc-constraint}

Among the DSB models treated in this paper, IYIT (Sec.~\ref{sec:IYIT}) relies on the
quantum-deformed constraint of $N_f=N_c$ SQCD \cite{Seiberg:1994bz}
\begin{equation}
  \det M-B\tilde B=\Lambda^{2N_c} .
  \label{eq:constraint}
\end{equation}
The phase of the invariant constraint can be determined from the symmetries of the
moduli space and compared with the phases of the pure-SYM condensates in
Sec.~\ref{sec:gaugino}.

\subsection{The phase of the right-hand side}
\label{sec:Nc-preview}

The bare angle $\theta_{{\rm bare},i}$ of Eq.~(\ref{eq:setup}) enters through
the holomorphic combination $\Lambda^b$ (Eq.~(\ref{eq:Lambda-b-phase})), with
$b=2N_c$ for $N_f=N_c$. This combination is positive at $\theta=0$ and
negative at $\theta=\pi$. A $k$-th root of a holomorphic quantity also carries
branch phases $e^{2\pi ij/k}$. These label isolated condensate vacua for
$S=(\Lambda^b)^{1/N_c}$ in pure SYM and local sheets at fixed $M$ for the ADS
superpotential (Sec.~\ref{sec:ADS}). The right-hand side of
Eq.~(\ref{eq:constraint}) contains the first power of $\Lambda^b$. For IYIT
($SU(2)$, $N_f=N_c=2$) it reads
$\Pf\hat M=\Lambda^4$ in terms of the antisymmetric $SU(4)_F$ meson $\hat M$ of
Sec.~\ref{sec:ARV}, where we establish the reality and normalization of the
numerical coefficient.
On the massive mesonic branch the branch label cancels in $\det M$, which
retains the sign of $\Lambda^{2N_c}$.

\subsection{Direct instanton computation at $m=0$ (Amati-Rossi-Veneziano)}
\label{sec:ARV}

Holomorphy, the flavor symmetries and the classical limit fix the form of the quantum
constraint to $\det M-B\tilde B=c\Lambda^{2N_c}$ \cite{Seiberg:1994bz}. The anomalous
axial charge of $\Lambda^{2N_c}$ matches that of $\det M$ and $B\tilde B$, so the scale
enters linearly and $c$ is a pure number. The $e^{i\theta}$ factor is included in
$\Lambda^{2N_c}$ by Eq.~(\ref{eq:Lambda-b-phase}). At $\theta=0$ and $\theta=\pi$,
$\Lambda^{2N_c}$ is real and nonzero, and CP acts by complex conjugation on the constraint
in the CP basis. Invariance of the constraint surface gives $c=c^*$. On the massive mesonic
branch $B=\tilde B=0$, holomorphic matching gives $\det M=\Lambda^{2N_c}$ in our scale
convention and fixes $c=1$ (Secs.~\ref{sec:mass-def} and \ref{sec:cancel}). The direct
instanton calculation below provides an independent check at zero mass.

The one-instanton origin of the right-hand side of Eq.~(\ref{eq:constraint}) was first
exhibited by the instanton computation of Amati, Rossi and Veneziano
\cite{Amati:1984uz}, whose selection rule, zero modes and squark profiles we use throughout
this subsection. This computation
is suited to examining any hidden phase, because it is carried out in a completely
massless theory with neither Higgsing nor a mass deformation, so there is
no concern about the instanton number fractionalizing somewhere along a
Higgsing chain --- it is, from beginning to end, a pure $\nu=1$ instanton of the unbroken
$SU(N_c)$ gauge theory. What it isolates is the structural origin of the phase: the pure-SYM
amplitude fixes only $\vevx{\lambda\lambda}^{N_c}$, so that the phase of the condensate enters
through an $N_c$-th root, while the $N_f=N_c$ amplitude contains $\Lambda^{2N_c}$ linearly, with
no root taken (the value of its coefficient was obtained in \cite{Finnell:1995dr}).

\paragraph{Method: the constant Green's function}
Consider the gauge-invariant Green's function built from the lowest components of chiral
superfields,
$G(x_1,\dots,x_n)=\vevx{0|T(A_1(x_1)\cdots A_n(x_n))|0}$
($A_i\in\{\lambda\lambda,\ \tilde Q^iQ_j\}$). From $\bar Q_{\dot\alpha}|0\rangle=0$ and
$[\bar Q_{\dot\alpha},A_i]=0$, the standard SUSY Ward identity gives
$\partial G/\partial x_i^\mu=0$, so $G$ is a constant independent of the $x_i$. It can
therefore be computed equally well at short relative distance (where the path
integral is dominated by a single weakly-coupled semiclassical instanton) or at
long relative distance (where, inserting a complete set of states between adjacent
operators, only the zero-energy states survive; since local operators do not induce
transitions between the possibly degenerate SUSY vacua $|\Omega\rangle$, this gives
$G=\sum_\Omega w_\Omega\vevx{\Omega|A_1|\Omega}\cdots\vevx{\Omega|A_n|\Omega}$, with
$w_\Omega=|\vevx{\Omega|0}|^2$ and $\sum_\Omega w_\Omega=1$). Equating these two evaluations
relates the instanton amplitude to the vacuum average of the products of the expectation
values $\vevx{\Omega|A_i|\Omega}$ above.

\paragraph{Selection rule}
Three conditions fix the numbers $n_\lambda$ and $n_\varphi$ of $\lambda\lambda$ and
$\tilde\varphi\varphi$ insertions in the $\nu$-instanton sector. Two of
them are $U(1)_R$ (the $R$-symmetry with $R(Q)=R(\tilde Q)=0$, anomalous for
$N_f\neq N_c$) and the anomalous $U(1)_A$ (the common phase rotation of $Q$ and $\tilde Q$): the
$2N_c\nu$ gluino zero modes and the $N_f\nu$ zero modes each of $\psi$ and $\tilde\psi$
violate these charges, and the insertions must carry exactly the violated charges. The third
is that the mass dimension of the constant $G$ be supplied by $\Lambda^{b\nu}$ with
$b=3N_c-N_f$. Together they exclude $\tilde\psi\psi$ and field-strength insertions and give
(App.~\ref{app:selection})
\begin{equation}
  n_\lambda=(N_c-N_f)\,\nu,\qquad n_\varphi=N_f\,\nu .
  \label{eq:selection}
\end{equation}
For $\nu=1$, $N_f<N_c$ gives the familiar ADS-type Green's
function. For $N_f=N_c$, $n_\lambda=0$, and the minimal constant Green's function
contains no $\lambda\lambda$ insertion at all, being built purely from the $N_c$ meson
bilinears $\tilde Q^iQ_i$. This is the crux of the argument below.

\paragraph{Zero modes}
The $\nu=1$ $SU(N_c)$ instanton has $2N_c$ explicit gluino zero modes in closed form,
$\lambda_0^{r,R}(x)$ ($r=1,\dots,N_c$, $R=1,2$). Four of these ($r=1,2$,
$R=1,2$) are the supersymmetric and superconformal modes of the $SU(2)\subset SU(N_c)$ in which
the instanton is embedded; the remaining $2(N_c-2)$ ($r=3,\dots,N_c$, $R=1,2$) have the
same doublet profile $\rho f(x)^{3/2}$ as the quark zero modes $\psi_0,\tilde\psi_0$, one
per flavor, with $f(x)=((x-a)^2+\rho^2)^{-1}$, where $a_\mu$ and $\rho$ are the center and
size of the instanton. Solving the
classical squark equation of motion $D^2\varphi_0=-i\sqrt2\,\lambda_0\psi_0$ in the
instanton background gives an explicit classical squark profile, sourced by the fermion zero modes and built from the same
collective coordinates as $\lambda_0,\psi_0$. As a result, for $N_f=N_c$, the $2N_c$ gluino
zero modes and $2N_c$ quark zero modes required to saturate the Grassmann
collective-coordinate integral are entirely absorbed by the $N_c$ squark profiles
that make up $\tilde\varphi\varphi$ --- the operator insertion $\lambda\lambda$ never
appears.

\paragraph{Pure SYM: a real integral and the origin of the $N_c$-fold phase}
For comparison, we first apply this computation to pure SYM. Consider
$G_{N_c}^0=\vevx{0|\lambda\lambda(x_1)\cdots\lambda\lambda(x_{N_c})|0}$. Pairing the
$2(N_c-2)$ zero modes with $r\ge3$ into $N_c-2$ pairs
($\lambda_0^{r,R}\lambda_0^{s,S}\propto\epsilon^{RS}f(x)^3\rho^2\delta^{rs}$) and the
remaining $r,s=1,2$ between two points
($\propto f(x_i)^4f(x_j)^4\rho^6(x_i-x_j)^2$), and carrying out the Grassmann integral and the
bosonic collective-coordinate integral ($a_\mu,\rho$) exactly using Feynman parameters
$\alpha_i$ ($\sum_i\alpha_i=1$), one obtains
\begin{equation}
  G_{N_c}^0=\gamma_{N_c}^0\,\Lambda^{3N_c} .
  \label{eq:GN0}
\end{equation}
$\gamma_{N_c}^0$ is real. At $\theta=0$, the instanton weight, the
collective-coordinate measure and the Feynman-parametrized position-space
integral are positive, and the Grassmann integral is real\footnote{%
The last of these is the only one whose positivity is not immediate. Carrying the
Feynman-parameter integral out in closed form gives
$\int d^4a\,d\rho^2\,(\rho^2)^{3N_c-2}\sum_{i<j}(x_i-x_j)^2f_i^4f_j^4\prod_{k\neq i,j}f_k^3
=\pi^2/[9(3N_c-1)]$, which is positive, and which is independent of the $x_i$ even though
the integrand is not, as the constancy of $G$ requires. For $N_c=2$ it reproduces the value
of \cite{Shifman:1999mv}. The remaining factors are the group volume,
which is positive, and the Grassmann combinatorics, which is real with a sign fixed by the
convention for the bilinear $\lambda\lambda$. Amati, Rossi and Veneziano evaluate
Eq.~(\ref{eq:GN0}) only up to an overall multiplicative constant.}.
For a unit-charge instanton the entire angle dependence is the factor $e^{i\theta}$
included in $\Lambda^{3N_c}$.

By clustering, $G_{N_c}^0=\vevx{\lambda\lambda}^{N_c}$, so Eq.~(\ref{eq:GN0}) fixes the
real, $Z_{2N_c}$-invariant combination $\vevx{\lambda\lambda}^{N_c}$ and nothing
else\footnote{%
The magnitude of $\vevx{\lambda\lambda}$ obtained this way needs to be corrected. This
computation, carried out without a Higgs VEV, is known as the ``strong-coupling instanton''
computation; for $SU(2)$ it gives $\vevx{S}^2=\tfrac45\Lambda^6$, whereas holomorphic
decoupling from the weakly-coupled Higgs phase gives $\Lambda^6$
\cite{Finnell:1995dr,Shifman:1999mv}. The cluster decomposition used to extract
$\vevx{\lambda\lambda}$ from this computation has been argued to fail
\cite{Hollowood:1999qn}, and a computation on $\mathbb{R}^3\times S^1$ reproduces the
weak-coupling value \cite{Davies:1999uw}. For $N_f=N_c$, in
contrast, the same computation reproduces the exact coefficient of
Eq.~(\ref{eq:constraint}) \cite{Finnell:1995dr} (see below Eq.~(\ref{eq:detM-instanton})).
Only the reality of $\gamma_{N_c}^0$ is used here.}. This
instanton computation is invariant under the microscopic (anomaly-free)
$Z_{2N_c}\subset U(1)_R$ symmetry by construction, and gives no reason to select any
particular phase. Consistency with the spontaneous breaking $Z_{2N_c}\to Z_2$ induced by
$\vevx{\lambda\lambda}\neq0$ therefore requires the form
\begin{equation}
  \vevx{\lambda\lambda}_k=(\gamma_{N_c}^0)^{1/N_c}|\Lambda|^3
  \exp\!\left[\frac{i(\theta+2\pi k)}{N_c}\right],
  \qquad k=0,\dots,N_c-1 ,
  \label{eq:lamlam-branch}
\end{equation}
but this is inserted after the fact by the representation theory of the unbroken
discrete symmetry, and is not itself directly computed. The raw instanton integral produces
only the real, $Z_{2N_c}$-invariant combination $\vevx{\lambda\lambda}^{N_c}$. This is the
precise, first-principles version of the phase discussed in Sec.~\ref{sec:VY}, and it
locates the phase ambiguity not in the instanton computation itself but in the
additional step of taking the $N_c$-th root of the invariant to extract the VEV of
a single field.

\paragraph{$N_f=N_c$ SQCD: a real coefficient with no root}
We now carry out the same computation for $N_f=N_c$. Since $n_\lambda=0$ here, the
relevant Green's function is
\begin{equation}
  \bigl(G_{N_c}^{N_c}\bigr)^{i_1\cdots i_{N_c}}_{j_1\cdots j_{N_c}}
  =\vevx{0|\tilde\varphi^{i_1}\varphi_{j_1}(x_1)\cdots
  \tilde\varphi^{i_{N_c}}\varphi_{j_{N_c}}(x_{N_c})|0} .
\end{equation}
Carrying out the same Feynman-parametrized collective-coordinate integral, now with the
squark profiles built from the gluino and quark zero modes, one again
obtains a real coefficient. The squark bilinear $\tilde\varphi_0\varphi_0$ is nonvanishing
for exactly the same pairs of gluino zero modes as $\lambda_0\lambda_0$, and its ratio to
$\lambda_0\lambda_0$ is a positive function of $x$, proportional to $1/(\rho^2f(x))$ with a
coefficient that differs between the pairs with $r,s\le2$ and those involving $r\ge3$.\footnote{%
Amati, Rossi and Veneziano quote this ratio as the constant $1/6$; the explicit
solution of the squark equation used here gives instead the $x$-dependent ratio
$1/(\rho^2f(x))$ times a coefficient that depends only on which zero modes are paired.
This is what the ancillary script described below checks. The position independence of
the final result holds in either case.} The
position independence therefore follows from carrying out the Grassmann and
collective-coordinate integrals directly. The Grassmann integral over the $2N_c$ gluino
collective coordinates gives a real constant times
$\rho^2\sum_{i<j}(x_i-x_j)^2f_i^3f_j^3\prod_{k\neq i,j}f_k^2$ (the quark and antiquark
collective coordinates appear once per flavor and integrate trivially), and after Feynman
parametrization the dependence on the positions drops out as in Eq.~(\ref{eq:GN0}):
$\int d^4a\,d\rho^2\,(\rho^2)^{2N_c-2}\sum_{i<j}(x_i-x_j)^2f_i^3f_j^3\prod_{k\neq i,j}f_k^2
=\pi^2/[4(2N_c-1)]$, independent of the $x_i$.\footnote{%
  A Python script that carries out this computation explicitly (zero modes, squark
  profiles, Grassmann and collective-coordinate integrals), together with the $SU(4)_F$
  averages used below, is provided as an ancillary file with the arXiv version of this paper.
  The zero modes, squark profiles and the Grassmann integral are checked there for
  $N_c=2,3$, and the collective-coordinate integral for $N_c=2,3,4$.}
By mass dimension the power of $\Lambda$ drops from $3N_c$ to $2N_c$:
\begin{equation}
  \bigl(G_{N_c}^{N_c}\bigr)^{i_1\cdots i_{N_c}}_{j_1\cdots j_{N_c}}
  =\epsilon^{i_1\cdots i_{N_c}}\epsilon_{j_1\cdots j_{N_c}}\;\gamma\,\Lambda^{2N_c},
  \qquad \gamma\ \text{real}.
\end{equation}
In the clustering limit $|x_i-x_j|\to\infty$, the left-hand side becomes the weighted
average over the vacua, so what the instanton fixes is the vacuum average
$\vevx{\det M}\equiv\sum_\Omega w_\Omega\det M(\Omega)
=N_c!\,\gamma\,\Lambda^{2N_c}$ over the degenerate vacua related by the flavor symmetry.
The $B\tilde B$ term contributes to the average as well: $\vevx{\tilde B(x_1)B(x_2)}$ is a
constant Green's function of the same $\nu=1$ sector and is nonvanishing at $m=0$
\cite{Amati:1984uz}. For $N_c=2$, the case used in IYIT
(Sec.~\ref{sec:IYIT}), the flavor symmetry is enlarged to $SU(4)_F$, under which $M^i_j$,
$B$ and $\tilde B$ combine into the antisymmetric meson $\hat M^{ab}=Q^a\!\cdot\!Q^b$
($a,b=1,\dots,4$; the $M^{ij}$ of Sec.~\ref{sec:IYIT}) with $\det M-B\tilde B=\Pf\hat M$.
Since the one-instanton measure is invariant under the anomaly-free $SU(4)_F$,
\begin{equation}
  \vevx{0|\hat M^{ab}(x_1)\,\hat M^{cd}(x_2)|0}=\gamma\,\epsilon_{abcd}\,\Lambda^4
\end{equation}
with the same $\gamma$, and this gives (App.~\ref{app:SU4})
\begin{equation}
  \vevx{\det M}=2\gamma\Lambda^4,\qquad \vevx{B\tilde B}=-\gamma\Lambda^4 .
\end{equation}
The instanton computation thus determines the vacuum average
\begin{equation}
  \vevx{\det M}-\vevx{B\tilde B}=3\gamma\,\Lambda^4\qquad(N_c=2),
  \label{eq:detM-instanton}
\end{equation}
with $\gamma$ real and $\Lambda^{2N_c}$ entering linearly. The combination
$\det M-B\tilde B$ takes the same value at every point of the moduli space, so the
normalized vacuum average gives $\det M-B\tilde B=3\gamma\Lambda^4$ there.
The value of $\gamma$ depends on the convention for
$\Lambda$; Finnell and Pouliot \cite{Finnell:1995dr} carried out this computation in the
$\overline{\rm DR}$ convention of Sec.~\ref{sec:ADS}, obtaining $\gamma=1/3$ for $N_c=2$
together with the ratio $\vevx{\det M}:\vevx{B\tilde B}=2:-1$ above, so that the coefficient in
Eq.~(\ref{eq:constraint}) is $1$. The sign of $\gamma$ is fixed by conventions
(the sign of the Yukawa coupling and the orderings of the spinor contractions and of the
Grassmann measure), and all that is needed here is that $\gamma$ is real.

\subsection{Mass deformation and the $N_c$-fold branch}
\label{sec:mass-def}

By the selection rule of the anomalous flavor symmetry together with holomorphy, the meson
VEV of massive $SU(N_c)$ SQCD ($N_f\le N_c$, diagonal mass
$m^{\tilde i}_i=m_i\delta^{\tilde i}_i$, real and positive for the discussion of branch
phases below) is fixed uniquely up to a computable numerical
coefficient \cite{Seiberg:1994bz,Intriligator:1995au}:
\begin{equation}
  M^i_{\tilde i}=\vevx{Q^i\tilde Q_{\tilde i}}
  =\bigl(\Lambda^{3N_c-N_f}\det m\bigr)^{1/N_c}
     \left(\frac1m\right)^i_{\tilde i} .
  \label{eq:M-mass}
\end{equation}
The root is taken of the full holomorphic product. Its $N_c$ determinations have
phases $(\theta+\arg\det m+2\pi k)/N_c$, with $k=0,1,\dots,N_c-1$.
For $N_f=N_c$ the relation follows directly from matching to pure SYM
(Sec.~\ref{sec:VY}). The common branch factor cancels in $\det M$,
leaving the phase of $\Lambda^{2N_c}$ (Sec.~\ref{sec:cancel}).

\paragraph{Step 1: flavor decoupling, no Higgsing}
The mass term $W_{\rm tree}=m_i\,Q^i\tilde Q_i$ is merely a deformation of the
superpotential; the squarks keep zero VEV and $SU(N_c)_{\rm gauge}$ remains unbroken.
Below $\mu_0=\min_im_i$, all $N_f=N_c$ quarks are integrated out and the theory flows,
with the gauge group entirely intact, to pure $SU(N_c)$ super Yang-Mills. The holomorphic
threshold matching relation is (with $b=3N_c-N_f=2N_c$)
\begin{equation}
  \Lambda_{\rm low}^{3N_c}=(\det m)\,\Lambda^{2N_c} .
  \label{eq:matching}
\end{equation}
No Higgsing occurs at this step --- the squarks decouple exactly as in an
ordinary QCD mass threshold, so $\pi_3(SU(N_c))=\mathbb{Z}$ is never disturbed. There is no
fractional instanton number and no discrete gauge symmetry acting along the way. The
concern about fractionalization of the instanton number is specific to Higgsing patterns
that leave an unbroken subgroup or an unbroken center, such as Higgsing in the adjoint
representation; here the matter is in the fundamental representation and the flavor mass
term breaks no gauge symmetry at all. This step is RG running across the matter
threshold.

\paragraph{Step 2: the same instanton as Sec.~\ref{sec:VY}}
In the deep IR, pure $SU(N_c)$ SYM confines and the gaugino condenses through
Eq.~(\ref{eq:WVY}). Its $N_c$ extrema are
\begin{equation}
  \vevx{S}_k=|\Lambda_{\rm low}|^3
  \exp\!\left[\frac{i(\theta+\arg\det m+2\pi k)}{N_c}\right],
  \qquad k=0,1,\dots,N_c-1 .
  \label{eq:S-branch}
\end{equation}
This is the instanton computation of Sec.~\ref{sec:VY}, for the same pure $SU(N_c)$
theory that the massive SQCD flows down to. No new source of phase appears anywhere, and the $N_c$-fold ambiguity is precisely the
Witten index of pure SYM itself.

\paragraph{Step 3: the meson VEV}
Treating $m_i$ as a source for $M^i_i=Q^i\tilde Q_i$ and substituting
Eq.~(\ref{eq:matching}) into $W_{VY}$, the effective superpotential is
\begin{equation}
  W_{\rm eff}(S,m)=S\left[\ln\frac{(\det m)\Lambda^{2N_c}}{S^{N_c}}+N_c\right]
\end{equation}
(in the same normalization as Eq.~(\ref{eq:WVY})). This extremum reproduces
Eq.~(\ref{eq:S-branch}). Since $m_i$ is the source of $M^i_i$,
$\vevx{M^i_i}=\partial W_{\rm eff}/\partial m_i$ \cite{Intriligator:1995au},
which agrees with the $\vevx{M^i_i}$ derived independently through Seiberg's
Lagrange-multiplier method $W=X(\det M-B\tilde B-\Lambda^{2N_c})+m_iM^i_i$
\cite{Seiberg:1994bz}, whose $B,\tilde B$ equations set $B=\tilde B=0$,
i.e.\ with Eq.~(\ref{eq:M-mass}). Using $\partial\ln\det m/\partial m_i=1/m_i$,
one finds
\begin{equation}
  M^i_i=\frac{\partial W_{\rm eff}}{\partial m_i}
  =\frac{\vevx{S}}{m_i}
  =\frac{|\Lambda^{2N_c}\det m|^{1/N_c}}{m_i}
   \exp\!\left[\frac{i(\theta+\arg\det m+2\pi k)}{N_c}\right] ,
  \label{eq:M-from-S}
\end{equation}
which reproduces Eq.~(\ref{eq:M-mass}) for $N_f=N_c$.

This identifies the $N_c$-fold branch of the meson VEV with the gaugino-condensation phase
of Sec.~\ref{sec:VY}, transmitted to every meson component through the same relation, with
the same integer $k$ appearing for every flavor index $i$ --- since there is only a single
$SU(N_c)$ sector confining in the deep IR, there is only a single choice of $k$, shared by
the entire matrix $M$. For positive $m_i$, the real mesonic branches are
precisely the real condensate branches identified in Sec.~\ref{sec:VY}.
The other branches come in complex-conjugate pairs and correspond to
spontaneous CP violation. The CP-violating branches of pure SYM thus reappear, with an identified microscopic
origin, whenever $N_f=N_c$ SQCD is moved away from the massless point.

\subsection{Cancellation of the branch in the invariant constraint}
\label{sec:cancel}

At $m\neq0$, where $M$ carries the branch of Eq.~(\ref{eq:M-from-S}), the branch
dependence cancels identically in the invariant $\det M$ that Eq.~(\ref{eq:constraint})
constrains. Using the precise result Eq.~(\ref{eq:M-from-S}), one can
see exactly why. Because the same $k$ multiplies every diagonal element $M^i_i$,
\begin{equation}
  \det M\ \propto\ \prod_{i=1}^{N_c}\left(e^{2\pi ik/N_c}\right)=e^{2\pi ik}=1
  \qquad\text{identically, for any }k .
\end{equation}
Every $k$ gives the same $\det M$, identically: this is a branch-independent algebraic
identity, because $\det M$ is built by combining
$N_c$ copies of the condensate of one and the same confining sector.

For general $N_c$, from Eq.~(\ref{eq:M-mass}),
\begin{equation}
  \det M=\frac{\vevx{S}_k^{N_c}}{\det m}
  =\frac{\Lambda^{2N_c}\det m}{\det m}=\Lambda^{2N_c} ,
  \label{eq:detM-cancel}
\end{equation}
with no trace of $k$ remaining. The constraint depends only on $\Lambda^{2N_c}$,
whose only $\theta$-dependence is through Eq.~(\ref{eq:Lambda-b-phase}); there is no
additional discrete phase from the instanton/matching computation itself. Seiberg
notes that this modification ``is due to a one instanton effect'' and that the relation
is ``true for every $m$ and not only in the limit $m\rightarrow0$''
\cite{Seiberg:1994bz} --- the cancellation found here is an exact identity holding across the entire $N_c$-sheeted family in
$m$.

Equations~(\ref{eq:detM-instanton}) and (\ref{eq:detM-cancel}) are independent computations,
and each yields a real result at either CP-invariant angle, with the sign of
$\Lambda^{2N_c}$ in the adopted normalization. The former is a direct computation at $m=0$ of the vacuum
average of the full combination $\det M-B\tilde B$ for $N_c=2$, in which $\Lambda^{2N_c}$
enters linearly, so no branch structure arises within the computation. The latter is evaluated at the
$m\neq0$ vacua, where $B=\tilde B=0$, and shows that the branches of the individual mesons
cancel when $\det M$ is assembled; it fixes the constant on the right-hand side of
Eq.~(\ref{eq:constraint}) and leaves the coefficient of $B\tilde B$ undetermined. For $N_c=2$
this coefficient is fixed by the anomaly-free $SU(4)_F$: every point $\hat M$ of the moduli
space can be brought to $B=\tilde B=0$ by an $SU(4)_F$ rotation, which leaves $\Pf\hat M$
invariant, and there $\Pf\hat M=\det M=\Lambda^4$ by Eq.~(\ref{eq:detM-cancel}). Hence
$\det M-B\tilde B=\Lambda^4$ at every point of the moduli space. For general $N_c$, the
coefficient of $B\tilde B$ is a pure number neutral under $U(1)_A$ and cannot depend on
$\Lambda$; the classical limit $\Lambda\to0$, in which $\det M=B\tilde B$ holds identically,
fixes it to $1$.

\paragraph{The massless limit}
The branch label $k$ belongs to $m\neq0$. There the mass term breaks
$SU(N_c)_L\times SU(N_c)_R$ explicitly, and the $N_c$ vacua $M_k=e^{2\pi ik/N_c}M_0$ are
isolated points related by a discrete symmetry. At $m=0$ the flavor symmetry is exact and
continuous, and $e^{2\pi ik/N_c}\mathbf{1}$ is an element of $SU(N_c)_L$ connected
continuously to the identity, so the $N_c$ points lie on a single orbit of the spontaneously
broken flavor symmetry, on the connected moduli surface of Eq.~(\ref{eq:constraint}).
At $m=0$, the case of IYIT, $M$ therefore carries no discrete branch; which point of the
surface is the vacuum is decided by the tree-level superpotential (Sec.~\ref{sec:Nc-IYIT}).

\paragraph{Comparison with pure SYM}
The $Z_{2N_c}$-invariant quantity computed by the instanton in pure SYM,
$\vevx{\lambda\lambda}^{N_c}$, is the $N_c$-th power of a single field, whose own
phase is a separate piece of data not fixed by the integral. $\det M$, in contrast, is not
the power of a single field: it is already a complete primitive
$SU(N_c)_L\times SU(N_c)_R$-invariant built from $N_c$ distinct meson components
$M^i_i$, and the instanton of Sec.~\ref{sec:ARV} computes its vacuum average all at once, as a whole --- there remains no
leftover single-field order parameter whose phase could be chosen independently.
Correspondingly, at the point $M^i_j=\Lambda^2\delta^i_j$, $B=\tilde B=0$ of the moduli
space, the unbroken symmetry is $SU(N_c)_V\times U(1)_B\times U(1)_R$
\cite{Seiberg:1994bz}; there is no leftover discrete subgroup of the kind that produced the
$Z_{2N_c}$ multiplicity in pure SYM.

\subsection{Implications for IYIT}
\label{sec:Nc-IYIT}

In IYIT, the quarks are strictly massless --- the only coupling is a tree-level Yukawa, not
a mass term --- so the theory always remains at $m=0$. The relevant question is therefore
not ``which of the $N_c$ discrete branches is realized'' (which only arises for $m\neq0$,
i.e.\ when the theory flows to pure SYM and confines), but rather ``where on the single surface
Eq.~(\ref{eq:constraint}), which contains a real slice, does the $F$-term potential built
from the Yukawa coupling select its minimum.'' This is the problem addressed in
Sec.~\ref{sec:IYIT}. Equation~(\ref{eq:constraint}) is a single-valued holomorphic
relation. Section~\ref{sec:IYIT-vacuum} shows that a real Yukawa coupling selects a vacuum
with a real meson VEV only in one of two sign classes; in the other the meson VEV is purely
imaginary and CP is spontaneously broken.

For check (i) of Sec.~\ref{sec:M3channels}, $\Lambda^{2N_c}$ enters the constraint
linearly (Sec.~\ref{sec:ARV}), and $\det M-B\tilde B$ takes the same value at every
point of the moduli space. For check (iii), the holomorphic matching introduces
no additional phase (Secs.~\ref{sec:mass-def} and \ref{sec:cancel}). The reality of
the selected vacuum, check (ii), is
checked in Sec.~\ref{sec:IYIT-vacuum} for the meson directions and in
Sec.~\ref{sec:IYIT-pseudo} for the pseudomodulus.

\section{IYIT SUSY breaking}
\label{sec:IYIT}

Following Izawa-Yanagida / Intriligator-Thomas \cite{Izawa:1996pk,Intriligator:1996pu}, we
take an $SU(2)$ gauge theory with four doublets $Q^i$ ($i=1,\dots,4$) and gauge-singlet
superfields $S^{ij}=-S^{ji}$:
\begin{equation}
  W=\sum_{i<j}\sum_{k<l}\lambda_{ij,kl}\,S^{ij}Q^k\!\cdot\!Q^l+\mc{A}\left(\Pf M-\Lambda^4\right),
  \label{eq:IYIT-W}
\end{equation}
where $Q^i\!\cdot\!Q^j\equiv\epsilon_{ab}Q_i^aQ_j^b$, $M^{ij}\equiv Q^i\!\cdot\!Q^j$ is the
gauge-invariant meson ($4\times4$ antisymmetric), and $\mc{A}$ is the Lagrange
multiplier enforcing the quantum constraint of Sec.~\ref{sec:Nc-constraint}.

\subsection{Reality of the couplings and the vacuum along the meson direction}
\label{sec:IYIT-vacuum}

Under the CP invariance of Eq.~(\ref{eq:setup}), $\lambda_{ij,kl}$ can be taken
real. The two CP-invariant angles give $\Lambda^4=+|\Lambda|^4$ at $\theta=0$
and $\Lambda^4=-|\Lambda|^4$ at $\theta=\pi$. The entire superpotential can be
taken real at either angle.

\paragraph{$SO(6)$ notation and the rank condition}
$M^{ij}$ is a rank-2 antisymmetric tensor of $SU(4)_F$, i.e.\ the representation
$\mathbf{6}$. Under $SU(4)\cong\mathrm{Spin}(6)$, $\mathbf6$ is the vector of $SO(6)$, whose
unique invariant quadratic form is the Pfaffian:
\begin{equation}
  \Pf M=\tfrac18\epsilon_{ijkl}M^{ij}M^{kl}
  =M^{12}M^{34}-M^{13}M^{24}+M^{14}M^{23}\equiv M\!\cdot\!M .
  \label{eq:IYIT-SO6}
\end{equation}
Below we write $M\in\mathbb{C}^6$ in components $M_I$ ($I=1,\dots,6$), and express the
quantum constraint as $M\!\cdot\!M=\Lambda^4$. The coupling is
$W=\lambda^{IJ}S_IM_J+\mc{A}(M\!\cdot\!M-\Lambda^4)$.

With $n$ singlets ($\lambda$ then an $n\times6$ matrix), setting $S=\mc{A}=0$ gives
$F_M=0$, so the tree-level potential is
\begin{equation}
  V=\sum_I\bigl|(\lambda M)_I\bigr|^2
  \qquad\text{on}\qquad M\!\cdot\!M=\Lambda^4 .
  \label{eq:IYIT-tree-pot}
\end{equation}
Let
\begin{equation}
  \ker\lambda\equiv\bigl\{M\in\mathbb{C}^6 \bigm| \lambda^{IJ}M_J=0\ \ \forall I\bigr\}
\end{equation}
be the subspace of meson directions on which the singlet couplings vanish. Then

\begin{itemize}
\item If $\ker\lambda$ contains even one element with $M\!\cdot\!M\neq0$, it can be
  rescaled to satisfy the constraint, giving $V=0$, i.e.\ SUSY is
  unbroken. SUSY breaking therefore requires $\ker\lambda$ to be totally isotropic
  ($\forall M\in\ker\lambda:\ M\!\cdot\!M=0$). Since the maximal isotropic subspace of a
  nondegenerate quadratic form on $\mathbb{C}^6$ has dimension 3, this requires
  $\mathrm{rank}\,\lambda\ge3$.
\item However, if $\ker\lambda$ is isotropic with $\dim\ker\lambda\ge1$, a runaway
  results. Splitting $M=M_\perp+M_0$ with $M_0\in\ker\lambda$ and $M_\perp$ in a complement of
  $\ker\lambda$, since $\Pf|_{\ker\lambda}\equiv0$ one has
  $M\!\cdot\!M=M_\perp\!\cdot\!M_\perp+2M_\perp\!\cdot\!M_0=\Lambda^4$, and since
  the quadratic form is nondegenerate one can take $|M_\perp|\to0$ with
  $|M_0|\sim\Lambda^4/|M_\perp|\to\infty$, satisfying the constraint while
  sending $V=|\lambda M_\perp|^2\to0$.
\end{itemize}

\noindent
Hence
\begin{equation}
  \text{a stable SUSY-breaking vacuum}\iff\ker\lambda=0\iff\mathrm{rank}\,\lambda=6 ,
  \label{eq:IYIT-rank6}
\end{equation}
so that six singlets are required, and $\lambda$ must be nondegenerate (even with
six singlets, if $\mathrm{rank}\,\lambda=5$ a SUSY vacuum generically reappears; what matters
is the rank, not the number). This is the essential reason IYIT introduces the whole
$\mathbf6$.

\paragraph{Real and purely imaginary vacua for real $\lambda$}
Instead of a single real vacuum, the IYIT sector with real couplings exhibits two distinct classes of vacua depending on the parameter $\lambda$, one of which spontaneously breaks CP. 
In the basis $M^{ij}$ where the CP transformation acts as complex conjugation (so that real couplings imply CP conservation), we can order the six independent components as $M=(M^{12}, M^{13}, M^{14}, M^{23}, M^{24}, M^{34})^T$. The quantum constraint then reads $\Pf M = M^T \eta M = \Lambda^4$, where $\eta$ is explicitly given by
\begin{equation}
	\eta = \frac{1}{2}
	\begin{pmatrix}
		0 & 0 & 0 & 0 & 0 & 1 \\
		0 & 0 & 0 & 0 & -1 & 0 \\
		0 & 0 & 0 & 1 & 0 & 0 \\
		0 & 0 & 1 & 0 & 0 & 0 \\
		0 & -1 & 0 & 0 & 0 & 0 \\
		1 & 0 & 0 & 0 & 0 & 0
	\end{pmatrix}.
	\label{eq:eta-matrix}
\end{equation}
This matrix consists of three $2\times2$ blocks on the anti-diagonal. Its normalized eigenvectors and corresponding eigenvalues are
\begin{equation}
	\begin{aligned}
		\tfrac{1}{\sqrt{2}}(M^{12} + M^{34}):\ &+\tfrac{1}{2}, \qquad &&\tfrac{1}{\sqrt{2}}(M^{12} - M^{34}):\ &-\tfrac{1}{2}, \\
		\tfrac{1}{\sqrt{2}}(M^{13} - M^{24}):\ &+\tfrac{1}{2}, \qquad &&\tfrac{1}{\sqrt{2}}(M^{13} + M^{24}):\ &-\tfrac{1}{2}, \\
		\tfrac{1}{\sqrt{2}}(M^{14} + M^{23}):\ &+\tfrac{1}{2}, \qquad &&\tfrac{1}{\sqrt{2}}(M^{14} - M^{23}):\ &-\tfrac{1}{2}.
	\end{aligned}
	\label{eq:eta-eigenbasis}
\end{equation}
Denoting these eigenmodes as $M_{a+}$ and $M_{a-}$ ($a=1,2,3$), the constraint is expressed as $\Pf M = \frac{1}{2}\sum_a (M_{a+}^2 - M_{a-}^2)$. This explicitly shows that $\eta$ is a real symmetric matrix of signature $(3,3)$. (Note that rotating this into a diagonal form $\sum_I M_I^2 = \Lambda^4$ requires multiplying the three $M_{a-}$ directions by $i$; under this complex transformation, CP no longer acts as simple complex conjugation.)

At $S=0$, the tree-level potential is given by\footnote{\label{fn:IYIT-Kahler}%
	The K\"ahler potential of the composite meson is not under control for $|M|\sim\Lambda^2$,
	but it drops out: at $S=0$ one has $F_M=0$, so the potential
	$V=K^{S_I\bar S_J}(\lambda M)_I\overline{(\lambda M)_J}$ is built from $F_S$ alone, and the
	$U(1)_R$ of Eq.~(\ref{eq:IYIT-Ranomaly}) forbids $S$-$M$ kinetic mixing. The strong dynamics
	sees $S$ only through the combination $\lambda S$ \cite{Chacko:1998si}, so the
	$M$-dependence it induces in $K^{S\bar S}$ is suppressed by $\lambda^2$ and is subleading as
	long as $\lambda$ is perturbative.}
\begin{equation}
	V = M^\dagger H M, \qquad H \equiv \lambda^T \lambda \quad (\text{real symmetric positive-definite}).
\end{equation}

To analyze the minimum, we define a real symmetric matrix $N \equiv H^{-1/2} \eta H^{-1/2}$. By Sylvester's law of inertia, $N$ has three positive and three negative eigenvalues, which we denote as $\nu_k$. Expanding $M = H^{-1/2} O w$ with a real orthogonal matrix $O \in SO(6,\mathbb{R})$ that diagonalizes $N$, the potential and the constraint become
\begin{equation}
	V = \sum_k |w_k|^2 , \qquad \sum_k \nu_k w_k^2 = \Lambda^4 .
\end{equation}
Here, $M$ is real (pure imaginary) if and only if $w$ is real (pure imaginary). Let $\nu_*$ be the eigenvalue of $N$ with the maximum absolute value. We then have the bound
\begin{equation}
	|\Lambda^4| = \Bigl| \sum_k \nu_k w_k^2 \Bigr| \le \sum_k |\nu_k| |w_k|^2 \le |\nu_*| V 
	\quad \Longrightarrow \quad V \ge \frac{|\Lambda^4|}{|\nu_*|} .
	\label{eq:IYIT-reality-bound}
\end{equation}
The equality requires that $w_k = 0$ for all components except those with $|\nu_k| = |\nu_*|$, and that all non-vanishing $\nu_k w_k^2$ have the same phase as $\Lambda^4$. When the largest positive and the largest negative eigenvalues of $N$ differ in magnitude, the nature of the vacuum depends only on the relative sign between the extremal eigenvalue $\nu_*$ and $\Lambda^4$:
\begin{itemize}
	\item If $\mathrm{sign}\,\nu_* = \mathrm{sign}\,\Lambda^4$, then $w_k^2 > 0$. The vacuum is \textbf{real} ($M_{\mathrm{vac}} \in \mathbb{R}^6$), and CP is preserved.
	\item If $\mathrm{sign}\,\nu_* = -\mathrm{sign}\,\Lambda^4$, then $w_k^2 < 0$. The vacuum is \textbf{pure imaginary} ($M_{\mathrm{vac}} = \pm i m$ for real $m$).\footnote{%
  Purely imaginary meson vacua also occur in softly broken massive SQCD at the
  CP-invariant angle $\bar\theta=\pi$ \cite{Csaki:2024lvk}. Here the
  quarks are massless and coupled to singlets.}
\end{itemize}
At fixed real Yukawa couplings, changing $\theta$ from zero to $\pi$
therefore exchanges the two classes.

When $+|\nu_*|$ and $-|\nu_*|$ are both eigenvalues of $N$, the extremal eigenvalue is degenerate in sign, the two cases above are not exhaustive, and the minimizing set is a degenerate family containing real, purely imaginary, and mixed points. This boundary is discussed below.

These two classes form open sets in the parameter space of real $\lambda$. An odd permutation of the flavors (e.g., $Q^1 \leftrightarrow Q^2$) acts as an orthogonal transformation $R$ satisfying $R^T \eta R = -\eta$, meaning that redefining $\lambda \to \lambda R$ maps one class into the other. For generic random couplings, both classes occur with roughly equal probability. 

In the pure imaginary vacuum, the $F$-term $F_S = -\overline{(\lambda M)}$ is also pure imaginary, so the rephasing invariant $\arg(\kappa F_X \overline{M}_{\mathrm{mess}}) = \pm \pi/2$ and the gluino mass acquires an $\mathcal{O}(1)$ phase. This vacuum is not CP-invariant: CP maps the meson VEV to its conjugate, $\pm i m \to \mp i m$. The anomaly-free $Z_4$ center of $SU(4)_F$ ($Q \to iQ$, $M \to -M$, $S \to -S$)\footnote{The center is anomaly-free with respect to the confining $SU(2)$: a rotation $Q \to e^{i\alpha}Q$ shifts the $\theta$ angle by $\Delta\theta = 2T(R)N_f\alpha = N_f\alpha$ (with $T(R)=\tfrac12$ for the doublet and $N_f=4$), and the center is the element $\alpha = 2\pi/N_f$, for which $\Delta\theta = 2\pi$; instantons therefore leave this discrete symmetry unbroken.} --- like CP itself --- relates the two CP-conjugate vacua and flips the sign of $F_S$ ($\pm \pi/2 \to \mp \pi/2$), but it cannot make the $F$-term real; the product of CP and this center leaves each of the two vacua invariant.
The $Z_4$ center enforces degeneracy in the isolated IYIT sector. With a fixed
messenger mass, the real messenger coupling breaks the center and makes the
phase observable in the gluino mass. If the effective IYIT--messenger
interactions preserve CP, the two purely imaginary vacua remain degenerate
as long as both survive.

In the real class, the two CP-invariant vacua $M$ and $-M$ are degenerate
in the isolated IYIT sector by the $Z_4$ center symmetry. The messenger
coupling breaks this symmetry and can lift their degeneracy.

\paragraph{The maximally symmetric point}
The two classes are separated by the measure-zero surface on which the largest positive and the largest negative eigenvalues of $N$ have equal magnitude; on this surface the minimum is degenerate and the minimizing set includes real, purely imaginary, and mixed points. The highly symmetric point $\lambda^{IJ} \propto \delta^{IJ}$, where $N \propto \eta$ and all three pairs of eigenvalues have equal magnitude, is the most symmetric point of this boundary. It is not generic, and the degeneracy can easily be broken by messenger loops; away from this surface the vacuum belongs to one of the two classes.

\subsection{Stabilization of the pseudomodulus and the phase direction}
\label{sec:IYIT-pseudo}

With six singlets and $\mathrm{rank}\,\lambda=6$, a tree-level flat direction remains in
$S$ at fixed $M=M_{\rm vac}$. In the CP basis of Eq.~(\ref{eq:eta-matrix}), a tangent
vector $t$ to the Pfaffian constraint satisfies $t^T\eta M_{\rm vac}=0$. Vanishing
$F$-terms along all five tangent directions therefore requires
$\lambda^TS\propto\eta M_{\rm vac}$. The normal component of the superpotential gradient
$\lambda^TS+2\mc{A}\eta M_{\rm vac}$ fixes the Lagrange multiplier $\mc{A}$, leaving a
one-complex-dimensional pseudomodulus $S\propto(\lambda^T)^{-1}\eta M_{\rm vac}$.

To compute its one-loop mass explicitly, we use the $\theta=0$ benchmark
with $\lambda^{IJ}=\lambda\delta^{IJ}$ and solve the constraint.\footnote{Here the constraint is written in the diagonal basis,
$\sum_I M_I^2=\Lambda^4$; relative to Eq.~(\ref{eq:IYIT-SO6}) this is reached by multiplying
the three negative-signature directions by $i$ and rescaling $\Lambda$.} Taking
$M_{\rm vac}=\Lambda^2e_1$ and writing the tangent directions as
$M_a$ ($a=2,\dots,6$), and substituting
$M_1=\sqrt{\Lambda^4-\sum_aM_a^2}\simeq\Lambda^2-\frac1{2\Lambda^2}\sum_aM_a^2$, one finds
\begin{equation}
  W_{\rm eff}=\lambda\Lambda^2X-\frac{\lambda}{2\Lambda^2}X\sum_aM_a^2
  +\lambda\sum_aS_aM_a,\qquad X\equiv S_1 ,
  \label{eq:IYIT-Weff-OR}
\end{equation}
which is precisely the minimal O'Raifeartaigh model: $X$ is the pseudomodulus and
$F_X=-\lambda\Lambda^2\neq0$ is the goldstino direction. In canonical normalization this
takes the form $W_{\rm eff}=fX-\tfrac h2X\sum_a\tilde M_a^2+m\sum_aS_a\tilde M_a$ with
$hf=m^2$ identically\footnote{%
At $X=0$ the scalar masses squared of $\tilde M_a$ are $m^2\pm hf=\{0,2m^2\}$. The five
massless ones are the Nambu-Goldstone bosons of $SU(4)_F\to Sp(4)$, which acquire
tree-level masses for generic $\lambda$ \cite{Chacko:1998si}.}.

\paragraph{$U(1)_R$ and the phase direction}
Before computing the Coleman-Weinberg potential, one can already argue that the phase
dependence is forbidden by symmetry. IYIT has an anomaly-free $U(1)_R$\footnote{Here
``anomaly-free'' means that the $SU(2)$ gauge anomaly $[U(1)_R\text{-}SU(2)^2]$ vanishes
(the anomalies $[U(1)_R]^3$ and the mixed $U(1)_R$-gravity anomaly are generically
nonzero). In flat space the selection rules of a global symmetry hold independently of these
anomalies, and the only fact used in the text is that gauge instantons do not break
$U(1)_R$, so none of this affects the argument below.}:
\begin{equation}
  R(Q)=0,\quad R(S)=2,\quad R(\mc{A})=2,\qquad
  2\cdot(+1)+4\cdot\tfrac12\cdot(-1)=0 ,
  \label{eq:IYIT-Ranomaly}
\end{equation}
where the two terms of the $SU(2)$ anomaly are the contributions of the gaugino and the
four quark doublets, respectively. Since both $\Lambda$ and $\Pf M$ are $R$-neutral, the quantum constraint is likewise
invariant, and $U(1)_R$ is exact both perturbatively and nonperturbatively. Since $X$ is the
only light field with $R=2$,
\begin{equation}
  V_{\rm eff}(X)=V_{\rm eff}(|X|)
  \label{eq:IYIT-VeffX}
\end{equation}
holds to all orders. $\arg X$ is the R-axion direction, and is not lifted regardless of the
details of the K\"ahler potential.

\paragraph{One-loop Coleman-Weinberg stabilization of $|X|$}
In the renormalizable O'Raifeartaigh approximation with canonical kinetic terms, Shih's
theorem \cite{Shih:2007av} gives a positive one-loop curvature at the origin for a
pseudomodulus coupled to fields with $R$-charges 0 and 2. This charge assignment follows
from $R(Q)=0$ and $R(S)=2$ for general $\lambda$. For the equal-coupling benchmark in
Eq.~(\ref{eq:IYIT-Weff-OR}), with $\mu\equiv hX$, the explicit one-loop calculation gives
$\mathrm{STr}\,\mc{M}^2=0$ and a potential $V_1$ independent of $\arg\mu$, in agreement
with the $U(1)_R$ argument above. In this benchmark $V_1(|\mu|)$ is monotonically
increasing and the curvature at the origin is
\begin{equation}
  m_X^2=\frac{5(\ln4-1)}{16\pi^2}\,h^2m^2\;>\;0 .
  \label{eq:IYIT-mX2}
\end{equation}
Canonically normalizing the meson as $\tilde M_a=M_a/\Lambda$ in
Eq.~(\ref{eq:IYIT-Weff-OR}) identifies $f=\lambda\Lambda^2$, $h=\lambda$ and
$m=\lambda\Lambda$, so that $hf=m^2$ as used above and
$m_X^2=5(\ln4-1)\lambda^4\Lambda^2/16\pi^2$, up to the normalization of the composite
meson. At the calculable minimum $\vevx{X}=0$, $U(1)_R$ is unbroken and no R-axion appears
\cite{Chacko:1998si} (the logarithmic growth of the potential for large $X$ was found in
\cite{Arkani-Hamed:1997lye}). The non-calculable K\"ahler terms from the strong dynamics are
smaller than the calculable one-loop contribution by $O(\lambda^2/16\pi^2)$
\cite{Chacko:1998si}, so they cannot reverse the sign of $m_X^2$ while $\lambda$ is
perturbative. A global minimum at $|X|\sim\Lambda/\lambda$, where the effective description
breaks down, cannot be excluded; there $U(1)_R$ would be spontaneously
broken. Equation~(\ref{eq:IYIT-VeffX}) holds at any $|X|$, so the resulting R-axion would be
exactly flat in global SUSY and its phase would be fixed by the real constant $\mc{C}$
through Eq.~(\ref{eq:IYIT-Raxion-pot}) below, as it is for the 3-2 model in
Sec.~\ref{sec:32model}; that minimum is real whatever the value of $\vevx{|X|}$.

\paragraph{$U(1)_R$ breaking and the R-axion direction}
The constant term $\mc{C}$, needed in supergravity for the (nearly)
vanishing cosmological constant, cannot be compensated by any $R$-charge of the IYIT light fields.
Added to the superpotential, it therefore breaks $U(1)_R$ from the point of view of the IYIT
sector itself. This breaking is distinct from the anomalous-plus-spontaneous breaking of
Sec.~\ref{sec:gaugino} induced by gaugino condensation, which supplies $\mc{C}$ here and
carries the branch structure of Eq.~(\ref{eq:C-def}). In what follows we take
$\mc{C}$ real, as for a separate pure $SU(2)$ condensate at its vanishing hidden angle.
We also work in the real IYIT class, in which the coefficient $f$ of $X$ is real.
For the benchmark above, $f=\lambda\Lambda^2$. The phase $\arg X$ is then
lifted by the potential
\begin{equation}
  V(X)=m_X^2|X|^2-\frac{4f|\mc{C}|}{M_P^2}\,|X|\cos(\arg X-\arg\mc{C}) .
  \label{eq:IYIT-Raxion-pot}
\end{equation}
Since $X$ is massive at the $R$-symmetric origin, $\mc{C}$ only shifts the minimum away
from it, to $\vevx{X}=2f\mc{C}/(m_X^2M_P^2)$ with $\arg\vevx{X}=\arg\mc{C}$, which is real
for real $\mc{C}$.

From this, the reality of $\vevx{W}$ and of the $F$-term follows directly as well. Before
adding the SUGRA constant term $\mc{C}$, $U(1)_R$ is not spontaneously broken
($\vevx{X}=0$), and since $R(W)=2\neq0$, $\vevx{W}=0$, which is trivially real. Adding
$\mc{C}$ breaks $U(1)_R$ explicitly, shifting $\vevx{X}=2f\mc{C}/(m_X^2M_P^2)$
to a real value (Eq.~(\ref{eq:IYIT-Raxion-pot})). Then
$\vevx{W}=\mc{C}+f\vevx{X}+O(\vevx{X}^2)=\mc{C}\bigl(1+2f^2/(m_X^2M_P^2)\bigr)$,
which remains real for real $f$ and $\mc{C}$.

\paragraph{Coupling to messengers}
If the messenger pair $f,\bar f$ has a mass $M_{\rm mess}+\kappa X$ (the elementary-singlet
portal of \cite{Murayama:2007fe}), the $B$-term of the
messengers comes from the auxiliary field of $X$, $F^X=-K^{X\bar\jmath}\overline{W_{\jmath}}$.
Unlike $X_1$ in Sec.~\ref{sec:32model}, $X$ here is an elementary gauge singlet, uncharged
under the confining $SU(2)$; it couples to the strongly-coupled meson sector only through the
perturbative Yukawa $\lambda$ (the uncontrolled K\"ahler potential flagged in footnote~\ref{fn:IYIT-Kahler}
concerns $M$, not $X$), so $K_{X\bar X}\simeq1$ is a good approximation and
$F^X\simeq-\lambda\Lambda^2=F_X$, matching the value used above. The induced gaugino mass is
then
\begin{equation}
  M_3\propto\frac{\kappa F_X}{M_{\rm mess}+\kappa\vevx{X}}.
\end{equation}
Since $\kappa,M_{\rm mess}$ are real by the CP invariance of Eq.~(\ref{eq:setup}), $F_X$ is
real as just established, and $\vevx X$ is real (established above), $M_3$ is real: the
messenger portal introduces no phase. Expanding for $\kappa\vevx X\ll M_{\rm mess}$ and using
the $\mc{C}$-induced part of the displacement (the only part that can carry a phase),
$\vevx X=2f\mc{C}/(m_X^2M_P^2)=c_Z\,16\pi^2\mc{C}/(\lambda^3 M_P^2)$ from
Eq.~(\ref{eq:IYIT-Raxion-pot}), with $c_Z=2/[5(\ln4-1)]=1.04$ in the notation of
Eq.~(\ref{eq:Zeff-shift}),
\begin{equation}
  M_3\propto\frac{\kappa F_X}{M_{\rm mess}}
  \left(1-c_Z\,\frac{16\pi^2}{\lambda^3}\,\frac{\kappa\,\mc{C}}{M_P^2 M_{\rm mess}}\right) ,
  \label{eq:IYIT-M3-mess}
\end{equation}
where only the second term could carry $\arg\mc{C}$. It is real here, since $\vevx X$ is
exactly real.


\section{ISS metastable SUSY breaking}
\label{sec:ISS}

ISS metastable SUSY breaking \cite{Intriligator:2006dd}, whose notation and results we use
throughout this section, operates in the free magnetic range
of $SU(N_c)$ SQCD,
\begin{equation}
	N_c+1\le N_f<\tfrac32N_c ,
\end{equation}
which requires $N_c\ge3$. The question addressed here is whether the ISS mechanism generates
a hidden phase of the kind seen in pure-SYM gaugino condensation.

\paragraph{Magnetic superpotential}
In the free magnetic range, the theory is described by a Seiberg-dual magnetic
$SU(\tilde N_c)$ gauge theory ($\tilde N_c\equiv N_f-N_c$), $N_f$ flavors of dual quarks
$q,\tilde q$, and a gauge-singlet meson $\Phi$ ($N_f\times N_f$), with tree-level
superpotential
\begin{equation}
	W=h\,\Tr(\Phi q\tilde q)-h\Tr(\mu^2\Phi) .
	\label{eq:ISS-W}
\end{equation}
Equation~(\ref{eq:ISS-W}) is a polynomial in $\Phi,q,\tilde q$ generated at tree
level in the weakly-coupled magnetic description; it takes no root of a nonperturbative
quantity and so carries no branch in the sense of Sec.~\ref{sec:Nc-preview}. That the
SUSY-breaking vacuum it selects is real for non-degenerate quark masses is shown below;
in the degenerate or nearly degenerate limit the vacuum orientation is instead a source of
spontaneous CP violation. Gauging the
magnetic $SU(\tilde N_c)$ adds a second, nonperturbative term that does carry a fractional
power; it is treated at the end of this section.

\paragraph{Phase of the duality map}
The electric theory has two CP-odd invariants once the messenger
portal $(\kappa_{ij}/M_*)\tilde Q^iQ^jf\bar f$ used below is included,
\begin{equation}
	\bar\theta_e\equiv\theta_e+\arg\det m
	\qquad\text{and}\qquad
	\arg\det(\kappa m^{-1}) ,
	\label{eq:ISS-thetabar-e}
\end{equation}
with $\theta_e$ the electric $\theta$-angle. The phases $\theta_e$, $\arg\det m$ and
$\arg\det\kappa$ separately are basis dependent: an anomalous flavor rotation shifts them
while leaving the two invariants of Eq.~(\ref{eq:ISS-thetabar-e}) fixed. Exact CP in the
UV sets both phases to $0$ or $\pi$ when the determinants are nonzero.
The second is a tree-level invariant of the explicit couplings.
The angle $\bar\theta_e$ is the phase of the
nonperturbative invariant $\Lambda^{3N_c-N_f}\det m$, and, as we will see, it enters the
superpotential Eq.~(\ref{eq:ISS-Wdyn-el}) that governs the supersymmetric vacua through a
fractional power, with a branch of the kind of Sec.~\ref{sec:Nc-preview}. These
invariants are not, however, the whole story: as we discuss below, even with both set to
$0$ or $\pi$ the vacuum orientation can spontaneously break CP when the quark-mass matrix
is degenerate or nearly so.

For a generic mass matrix $m$, we can pass to the basis where $m$ is real and diagonal with non-negative entries, $m = \diag(m_1, \dots, m_{N_f})$. The dual description consists of
\begin{equation}
	W_{\rm dual}=\frac{1}{\hat\Lambda}\Tr(Mq\tilde q)+\Tr(m M) ,
	\qquad
	K\supset\frac{1}{\alpha|\Lambda|^2}\Tr M^\dagger M ,
	\label{eq:ISS-dual-WK}
\end{equation}
where the coefficient $\hat\Lambda$ of the dual superpotential is constrained by the
scale-matching relation
\begin{equation}
	\Lambda^{3N_c-N_f}\tilde\Lambda^{3\tilde N_c-N_f}=(-1)^{N_f-N_c}\hat\Lambda^{N_f} ,
	\label{eq:ISS-matching}
\end{equation}
and the dictionary to the canonically normalized variables of Eq.~(\ref{eq:ISS-W}) is
\begin{equation}
	\Phi=\frac{M}{\sqrt\alpha\,|\Lambda|},\qquad
	h=\frac{\sqrt\alpha\,|\Lambda|}{\hat\Lambda},\qquad
	\mu^2=-m\,\hat\Lambda,\qquad
	\Lambda_m=\tilde\Lambda
	\label{eq:ISS-dictionary}
\end{equation}
We choose $\hat\Lambda>0$ by a phase redefinition of the magnetic quarks. Then $h>0$
and $\mu^2$ has real negative eigenvalues for positive quark masses. The phase of the
electric holomorphic scale is retained in Eq.~(\ref{eq:ISS-matching}). The physical
auxiliary field follows directly from
Eq.~(\ref{eq:ISS-dual-WK}),
\begin{equation}
	F^M=-K^{M\bar M}\,\overline{W_M}
	=-\alpha|\Lambda|^2\Bigl(m+\frac{q\tilde q}{\hat\Lambda}\Bigr)^{\!\dagger} .
	\label{eq:ISS-FM-general}
\end{equation}
Assuming $m$ is non-degenerate, at the metastable vacuum the rank-$\tilde N_c$ matrix $\vevx{q\tilde q}$ cancels the $\tilde N_c$ eigenvalues of $\mu^2 = -m\hat\Lambda$ that are largest in magnitude. $F^M$ vanishes on this Higgsed block, and on the remaining $N_c\times N_c$ block it aligns uniquely with the smallest masses,
\begin{equation}
	F^M=-\alpha\,|\Lambda|^2\,\diag(m_{i\in\text{remaining}}) .
	\label{eq:ISS-FM}
\end{equation}
Three quantities drop out here, and they are the three places where a phase could have entered.
First, $\hat\Lambda$ cancels between $\vevx{q\tilde q}$ and $1/\hat\Lambda$. It is in any
case not fixed by the electric theory: the rescaling $q\to cq$, $\tilde q\to c\tilde q$
sends $\hat\Lambda\to c^2\hat\Lambda$ and is anomalous under the magnetic gauge group,
so what the electric theory determines is the invariant
combination $q\tilde q/\hat\Lambda=-m$, and $\hat\Lambda$ itself is a normalization
convention rather than a physical branch. Second, the holomorphic scale enters only as
$|\Lambda|^2$ through the K\"ahler metric, and $\alpha|\Lambda|^2>0$ is real by the
Hermiticity and positivity of $K$ alone, with no appeal to CP. Third, $\bar\theta_e$ is
absent, because Eq.~(\ref{eq:ISS-FM}) takes no root of the invariant
$\Lambda^{3N_c-N_f}\det m$, which is the only quantity that carries it; the one term that
does take such a root is treated at the end of this section. Hence $F^M$ is real for real
$m$, and its sign is fixed as well. This is a statement in the basis in which the
Lagrangian is manifestly CP invariant: under the anomalous flavor rotation
$Q,\tilde Q\to e^{i\varphi}(Q,\tilde Q)$
that trades $\arg\det m$ for $\theta_e$, $M$ and $F^M$ pick up $e^{2i\varphi}$, $m$ picks up
$e^{-2i\varphi}$ and $|\Lambda|^2$ is invariant, so Eq.~(\ref{eq:ISS-FM-general}) is covariant
rather than invariant. The coupling $\kappa$ of the messenger portal
$(\kappa/M_*)\tilde QQf\bar f$ used below multiplies the same quark bilinear as $m$ and
rotates with it. The leading messenger insertion
$\operatorname{Tr}(\kappa^T F^M)$ is invariant under this rephasing
and independent of $\bar\theta_e$.
For fixed real $m$ and $\kappa$, switching the electric bare angle from
$\theta_e=0$ to $\theta_e=\pi$ leaves this metastable auxiliary field and
the leading messenger-loop orientation potential below unchanged.

In the canonically normalized variables, the remaining block has
$F_\Phi=(h\mu^2)^\dagger=-\sqrt\alpha\,|\Lambda|m$ for real diagonal $m$.
The relation $F^M=\sqrt\alpha\,|\Lambda|F_\Phi$ then reproduces
Eq.~(\ref{eq:ISS-FM}) on this block.

\paragraph{K\"ahler potential}
The leading K\"ahler potential that ISS itself exhibits near the origin in the free magnetic
range is
\begin{equation}
	K=\frac1\beta\Tr\bigl(q^\dagger q+\tilde q^\dagger\tilde q\bigr)
	+\frac{1}{\alpha|\Lambda|^2}\Tr M^\dagger M+\cdots,
	\qquad\alpha,\beta\in\mathbb{R}_{>0} ,
	\label{eq:ISS-Kahler}
\end{equation}
whose meson part is the one already used in Eq.~(\ref{eq:ISS-dual-WK}); in the canonical
variables of Eq.~(\ref{eq:ISS-dictionary}) the second term is $\Tr\Phi^\dagger\Phi$. Here
$\alpha,\beta$ are real and positive by positivity of the Hermitian bilinears, and
the dictionary Eq.~(\ref{eq:ISS-dictionary}) is written in the normalization $\beta=1$ that
ISS adopt there. Since the reality of the order parameter follows from
Eq.~(\ref{eq:ISS-FM}) alone, which involves only $\alpha$ and $m$, no convention for
$q,\tilde q$ need be chosen, and the convention-dependence of the higher terms represented
by ``$\cdots$'' in Eq.~(\ref{eq:ISS-Kahler}) does not matter. These higher terms are themselves
unknown, but ISS itself shows explicitly that, since the magnetic gauge group is IR-free in
the free magnetic range, they are suppressed by powers of $1/\Lambda$ near
$M=q=\tilde q=0$ and can be neglected. This metastable SUSY-breaking vacuum sits at
$q\tilde q=\diag(\mu^2\mathbf1_{\tilde N_c},0)$, $\Phi=0$ ($\mu\ll\Lambda$), precisely in the regime where the
higher corrections are suppressed (the true SUSY vacuum lies instead at a different, larger
field value $|\mu|\ll|h\Phi|\ll|\Lambda_m|$, and is not the
subject of the discussion here).

\paragraph{SUSY breaking and the reality of the vacuum}
$F_\Phi=-[h(q\tilde q-\mu^2)]^\dagger$ cannot vanish: the rank of $q\tilde q$ is at most
$\tilde N_c<N_f$, while $\mu^2$ has rank $N_f$ --- an O'Raifeartaigh-type rank
condition, structurally identical to the mechanism of Sec.~\ref{sec:IYIT} (there,
$\mathrm{rank}\,\lambda$ against the Pfaffian constraint; here, $\mathrm{rank}(q\tilde q)$
against $\mu^2$).

If $m$ is non-degenerate, the energy-minimizing configuration uniquely forces $q\tilde q$ to cancel the $\tilde N_c$ eigenvalues of $\mu^2$ that are largest in magnitude, yielding
\begin{equation}
	q\tilde q=\begin{pmatrix}\mu^2_{\rm large}&0\\0&0\end{pmatrix},
	\qquad
	\Phi=\begin{pmatrix}0&0\\0&\Phi_0\end{pmatrix} ,
\end{equation}
with $\Phi_0$ (an $N_c\times N_c$ block) a classical pseudomodulus. As established above,
$F^M$ is real by Eq.~(\ref{eq:ISS-FM}), and taking $h\in\mathbb R_{>0}$ and real negative eigenvalues for $\mu^2$ makes this extremum manifestly real. The one-loop Coleman-Weinberg
potential for $\Phi_0$ is a real function of the real mass-squared eigenvalues of the
fluctuation spectrum around this point, and is minimized at $\Phi_0=0$, as the explicit
one-loop computation of ISS (whose
coefficients are positive) shows, in accordance with Shih's theorem
\cite{Shih:2007av,Intriligator:2007py}; this minimum is unique, so there is no
multiplicity of complex vacua\footnote{%
	At two loops the origin of a pseudo-moduli space of this type can instead be destabilized
	\cite{Giveon:2008wp}. That computation treats the free magnetic phase with massless and
	massive flavors together, or with all flavors massive but hierarchical in mass, which naturally aligns with our requirement of non-degenerate masses. Were the origin displaced, the reality established above would be unaffected in
	any case, since $F^M$ is fixed by Eq.~(\ref{eq:ISS-FM}) alone and does not
	depend on where $\Phi_0$ sits.}.

The situation is markedly different in the degenerate limit $m_i = m_0$. Here, the unbroken global symmetry $U(N_f)_V$ makes the orientation of the Higgsing highly degenerate, spanning a complex Grassmannian orbit $Gr(\tilde N_c, N_f)$. The configuration where $q\tilde q$ is a real diagonal matrix is just one arbitrary point on this orbit, and its apparent reality is merely an artifact of a choice of basis. The true vacuum orientation is dynamically selected only when coupling to messengers, which can induce spontaneous CP violation as discussed below.

\paragraph{Magnetic gaugino condensation}
In the free magnetic range the magnetic $SU(\tilde N_c)$ theory with $N_f$ flavors is
infrared free: $N_f<\tfrac32N_c$ is the same condition as $N_f>3\tilde N_c$. At the
metastable vacuum the magnetic gauge group is moreover completely Higgsed, since
$\vevx{q\tilde q}$ has full rank in color, so no unbroken
non-abelian factor is left to condense. The superpotential there is Eq.~(\ref{eq:ISS-W})
together with the one-loop Coleman-Weinberg potential, and the $U(1)_R$ under which
$R(\Phi)=2$, $R(q)=R(\tilde q)=0$ is an accidental symmetry of the low-energy theory.

A fractional power appears in the opposite regime, where the magnetic quarks are massive
rather than condensed. Once $\vevx{h\Phi}$ is large enough, the quarks have vanishing VEV
and integrating them out leaves an unbroken pure $SU(\tilde N_c)$ super Yang-Mills with
$\Lambda_L^{3\tilde N_c}=h^{N_f}\det\Phi\;\Lambda_m^{3\tilde N_c-N_f}$, and its gaugino
condensation gives
\begin{equation}
	W_{\rm dyn}=\tilde N_c
	\left(h^{N_f}\Lambda_m^{-(N_f-3\tilde N_c)}\det\Phi\right)^{1/\tilde N_c} .
	\label{eq:ISS-Wdyn}
\end{equation}
This carries the power $1/\tilde N_c$ and with it a $\tilde N_c$-fold branch, generically
complex for $\tilde N_c\ge3$, which the free magnetic range does allow ($N_c=7$, $N_f=10$
gives $\tilde N_c=3$); it also breaks the accidental $U(1)_R$ explicitly. Eliminating
$h,\Lambda_m,\hat\Lambda$ with Eqs.~(\ref{eq:ISS-matching}) and (\ref{eq:ISS-dictionary})
turns Eq.~(\ref{eq:ISS-Wdyn}) into a purely electric expression,
\begin{equation}
	W_{\rm dyn}=\tilde N_c\left((-1)^{N_f-N_c}\,
	\frac{\det M}{\Lambda^{3N_c-N_f}}\right)^{1/\tilde N_c} ,
	\label{eq:ISS-Wdyn-el}
\end{equation}
in which $\tilde\Lambda$ and $\hat\Lambda$ have both cancelled. This is the term that
carries $\bar\theta_e$. Extremizing $W_{\rm dyn}+\Tr(m M)$ gives $M \propto m^{-1}$, reproducing the $N_c$
supersymmetric vacua of the electric theory,
\begin{equation}
	\vevx{M}=\bigl(\Lambda^{3N_c-N_f}\det m\bigr)^{1/N_c}m^{-1},
	\qquad
	\vevx{W}=N_c\bigl(\Lambda^{3N_c-N_f}\det m\bigr)^{1/N_c} ,
	\label{eq:ISS-susy-W}
\end{equation}
whose phase is $(\bar\theta_e+2\pi k)/N_c$. At fixed $m$, changing the
electric bare angle from zero to $\pi$ shifts this phase by $\pi/N_c$.
The dependence on $\bar\theta_e$ and on the branch enters these
supersymmetric vacua through Eq.~(\ref{eq:ISS-Wdyn-el}). Those
vacua sit at parametrically large field value,
\begin{equation}
	\vevx{h\Phi}=\Lambda_m\,\varepsilon^{2\tilde N_c/(N_f-\tilde N_c)}\,\mathbf1_{N_f},
	\qquad \varepsilon\equiv\frac{\bar\mu}{\Lambda_m},\qquad
	|\varepsilon|\sim\sqrt{\frac{\bar m}{|\Lambda|}} ,
	\label{eq:ISS-susy-vac}
\end{equation}
$\varepsilon$ being the parameter that controls the longevity of the metastable state (where $\bar m$ and $\bar\mu$ denote the characteristic scale of the masses).
At the lower edge of the range, $N_f=N_c+1$, the magnetic gauge
group is trivial, $\tilde N_c=1$, and Eq.~(\ref{eq:ISS-Wdyn-el}) reduces to
$-\det M/\Lambda^{2N_c-1}$, which is exactly the determinant term of the confined
description,
$W=(\tilde B^TMB-\det M)/\Lambda^{2N_c-1}+\Tr(mM)$. The root is trivial there and no branch
arises.

\paragraph{$U(1)_R$ breaking by the constant term}
The constant term $\mc{C}$ carries no $R$-charge that the light fields can
compensate, so it breaks the accidental $U(1)_R$ explicitly as well. At the metastable
vacuum, where the magnetic gauge group is completely Higgsed, it is the only source, and $\arg\Phi_0$ strictly ceases to be flat. Collecting the
terms of the supergravity potential that are linear in $\mc{C}$,
\begin{equation}
	\Delta V=\frac{\overline{\mc{C}}}{M_P^2}
	\Bigl[K^{i\bar\jmath}K_{\bar\jmath}W_i-3W\Bigr]+{\rm h.c.} ,
	\label{eq:ISS-linear-C}
\end{equation}
and evaluating them with Eq.~(\ref{eq:ISS-Kahler}), the cubic coupling $h\Tr(q\Phi\tilde q)$
cancels between the two pieces and only the linear term $-h\Tr(\mu^2\Phi)$ survives.
To estimate the displacement, we use the equal-mass ISS benchmark, with $m_{\Phi_0}^2$
the common one-loop Coleman--Weinberg mass of the pseudomodulus:
\begin{equation}
	V(\Phi_0)=m_{\Phi_0}^2\Tr\Phi_0^\dagger\Phi_0
	+\frac{2h\mu^2}{M_P^2}\bigl(\overline{\mc{C}}\,\Tr\Phi_0+{\rm h.c.}\bigr) .
	\label{eq:ISS-C-tadpole}
\end{equation}
The source is the trace alone, so only the diagonal entries are shifted, each by the same
amount, and writing $\Phi_0^{aa}=|\Phi_0^{aa}|e^{i\vartheta_a}$ the dependence on
$\vartheta_a$ is a single $\cos(\vartheta_a-\arg\mc{C})$. Since $\Phi_0$ is massive at the
$R$-symmetric origin, $\mc{C}$ only shifts the minimum away from it: with $h>0$ and
real negative eigenvalues for $\mu^2$, the minimum is at $\vartheta_a=\arg\mc{C}$, i.e.\
$\vevx{\Phi_0}=-2h\mu^2\mc{C}/(m_{\Phi_0}^2M_P^2)\,\mathbf1_{N_c}$, which is real for real
$\mc{C}$. With $m_{\Phi_0}^2=(\ln4-1)\tilde N_ch^4|\bar\mu|^2/8\pi^2$, read off from
ISS, the one-loop origin of $m_{\Phi_0}^2$
enhances the magnitude of each diagonal shift to $c_Z\,16\pi^2|m_{3/2}|/h^3$ with
$c_Z=1/[(\ln4-1)\tilde N_c]=2.6/\tilde N_c$, independently of $\mu$ and $\Lambda$.

The tadpole of Eq.~(\ref{eq:ISS-C-tadpole}) sources $\Tr\Phi$, so it shifts not only the
pseudomodulus $\Phi_0$ but also the Higgsed $\tilde N_c\times\tilde N_c$ block, whose
tree-level mass is $|h\bar\mu|$, by $|\delta\Phi|\sim |m_{3/2}|/h$. Both shifts are far below $|\bar\mu|$, so the magnetic gauge group stays completely Higgsed and
Eq.~(\ref{eq:ISS-Wdyn-el}) does not apply. By analogy with the case $N'=1$ of
Sec.~\ref{sec:ADS}, we expect the remaining single-instanton effects of the Higgsed magnetic
group to take no root.

\paragraph{Coupling to messengers}
Whether a new phase enters upon coupling to a messenger sector has been checked individually,
for the explicit example of \cite{Murayama:2007fe}, in which the composite meson
$\tilde Q^iQ^j$ is coupled to a messenger pair $f,\bar f$ through the portal
$(\kappa_{ij}/M_*)\tilde Q^iQ^jf\bar f$. This portal couples to the electric meson $M$
itself, so the quantity entering $M_3$ is governed by $F^M$.

If the mass matrix $m$ is non-degenerate, $F^M$ is uniquely fixed at the tree level to the real diagonal form of Eq.~(\ref{eq:ISS-FM}), and $M_3 \propto \Tr(\kappa^T F^M)$ is then strictly real regardless of the flavor structure of $\kappa$: even antisymmetric components of the portal coupling $\kappa$ decouple from the real diagonal $F^M$. The messenger loops rotate this orientation only when the loop correction exceeds the tree-level gap set by the mass splitting; in that case, which includes the exactly degenerate limit, a $\kappa$ with an antisymmetric part can select a complex orientation.

Conversely, if the masses are degenerate, $m_i = m_0$, the orientation of the projector $\Pi$ ($\Pi^2=\Pi$, $\Pi^\dagger=\Pi$, an $N_f\times N_f$ matrix of rank $N_c$) in $F^M = -\alpha|\Lambda|^2 m_0 \Pi$ is undetermined within the ISS sector itself. The degeneracy is expected to be lifted by the messenger loops. The portal induces a messenger $B$-term $F_{\rm eff}\equiv\Tr(\kappa^T F^M)/M_*$, and the scalar mass-squared eigenvalues $M_{\rm mess}^2\pm|F_{\rm eff}|$ generate the one-loop Coleman-Weinberg potential
\begin{equation}
	V_{\rm CW}\simeq-\frac{5N_{\rm mess}}{16\pi^2}\,|F_{\rm eff}|^2\ln\frac{\Lambda_{\rm UV}^2}{M_{\rm mess}^2},
	\label{eq:ISS-CW-orientation}
\end{equation}
valid for $|F_{\rm eff}|\ll M_{\rm mess}^2\ll\Lambda_{\rm UV}^2$. Here $N_{\rm mess}$ counts the messenger pairs, assumed to be $\mathbf5\oplus\bar{\mathbf5}$ of $SU(5)$, whose five components give the factor $5$. Since $|F_{\rm eff}|\propto|\Tr(\kappa^T\Pi)|$, the messenger loops favor the orientation that maximizes $|\Tr(\kappa^T \Pi)|$. An antisymmetric component of $\kappa$ can favor a complex projector. The equal-mass limit can therefore give an $\mathcal{O}(1)$ phase in the gluino mass. Whether the one-loop messenger logarithm is the whole effect (higher loops, finite parts and the orientation dependence of the magnetic instantons are not computed here) is left open. Thus, while the ISS mechanism is safe from spurious phases whenever the quark-mass splittings exceed the messenger-loop corrections, a nearly degenerate spectrum, with splittings below the messenger-loop corrections, can admit complex orientations. The minimizing projector remained real for symmetric $\kappa$ in the scan described below. This contrasts sharply with the IYIT models of Sec.~\ref{sec:IYIT}, where roughly half of the parameter space (unknown real couplings) leads to purely imaginary vacua. In the scan summarized in Table~\ref{tab:ISS-clo}, the fraction of samples with a complex minimum decreased as the mass hierarchy increased.

The criterion can be made quantitative. Dividing by $\alpha|\Lambda|^2$, the orientation
dependence is governed by the dimensionless functional
\begin{equation}
	v(\Pi)=\Tr[m^2\Pi]-c\,\bigl|\Tr(\kappa^T\Pi m)\bigr|^2,
	\qquad
	c=\frac{5N_{\rm mess}}{16\pi^2}\,\frac{\alpha|\Lambda|^2}{M_*^2}
	\ln\frac{\Lambda_{\rm UV}^2}{M_{\rm mess}^2},
	\label{eq:ISS-vPi}
\end{equation}
to be minimized over real and complex projectors $\Pi$ of rank $N_c$. Since $v$ is CP
invariant and the second term depends on $\Pi$ only through the modulus
$|\Tr(\kappa^T\Pi m)|$, the global minimum is either exactly real (phase $0$ or $\pi$) or
one of a CP-conjugate pair of complex projectors. In the exactly degenerate limit, where
the minimum is the $|\Tr(\kappa^T\Pi m)|$-maximizer for every $c$, this phase is
$\mathcal{O}(1)$. Direct
global minimization of Eq.~(\ref{eq:ISS-vPi}) (multiple random starts of a quasi-Newton
method) for $(N_c,N_f)=(3,4)$ and a geometric mass spectrum $m=\diag(1,r,r^2,r^3)$ gives
the distribution of Table~\ref{tab:ISS-clo}.\footnote{A Python script that carries out
this scan (the global minimization of Eq.~(\ref{eq:ISS-vPi}) over real and complex
projectors, the location of the complex windows on a logarithmic $c$ grid with geometric
bisection of the edges, and the symmetric-$\kappa$ check), together with its output, is
provided as an ancillary file with the arXiv version of this paper.} For a given $\kappa$ the complex window is
the range of $c$ over which the global minimum is a complex projector, with lower edge
$c_{\rm lo}$;
the scan covers $10^{-5}\lesssim c\lesssim2$, which contains the range allowed by
Eq.~(\ref{eq:ISS-c-bound}). In the exactly degenerate limit $r=1$ the tree term of
Eq.~(\ref{eq:ISS-vPi}) is flat, so for the $\kappa$ whose $|\Tr(\kappa^T\Pi m)|$-maximizer
is complex --- about a third of the sample --- the threshold sits at $c_{\rm lo}=0$;
lifting the degeneracy lowers
the fraction of complex vacua and raises $c_{\rm lo}$ monotonically, up to the scale of
the bound of Eq.~(\ref{eq:ISS-c-bound}) at $r\gtrsim10$, where only
a small fraction of the couplings remain dangerous. The size of $c$ is set by the
portal scale. By the dictionary Eq.~(\ref{eq:ISS-dictionary}), $|h\mu^2|=\sqrt\alpha\,m|\Lambda|$ and hence
\begin{equation}
	c=\frac{5N_{\rm mess}}{16\pi^2}\Bigl(\frac{\sqrt\alpha|\Lambda|}{M_*}\Bigr)^2
	\ln\frac{\Lambda_{\rm UV}^2}{M_{\rm mess}^2},
	\label{eq:ISS-c-size}
\end{equation}
with $\Lambda_{\rm UV}\sim|\Lambda|$ the cutoff of the
meson effective theory. The health of the messengers,
$|F_{\rm eff}|=\kappa\alpha|\Lambda|^2m/M_*\ll M_{\rm mess}^2$, bounds the logarithm by
$\ln(\Lambda_{\rm UV}^2/M_{\rm mess}^2)\lesssim\ln(M_*/\kappa\alpha m)$, and since the
portal is a higher-dimensional operator generated by physics above the strong scale,
$M_*\gtrsim\sqrt\alpha|\Lambda|$, the coefficient is bounded by
\begin{equation}
	c\lesssim\frac{5N_{\rm mess}}{16\pi^2}\,\ln\frac{|\Lambda|}{\kappa\sqrt\alpha\,m},
	\label{eq:ISS-c-bound}
\end{equation}
$\simeq0.2\,N_{\rm mess}$ for $|\Lambda|/m=10^3$ and $\kappa\sqrt\alpha\sim1$, decreasing as
$(\sqrt\alpha|\Lambda|/M_*)^2$ for a portal from a higher scale. The lower part of the
threshold range of Table~\ref{tab:ISS-clo} is therefore reached only for a portal
saturated at the strong scale, $M_*\sim\sqrt\alpha|\Lambda|$; for a portal from genuinely
higher scales, $c\ll c_{\rm lo}$ and the tree-level orientation survives.
For the 20 real symmetric choices of $\kappa$ tested at each scanned mass ratio, the
minimizing projector remained real throughout the scan. Complex minima were observed for
$\kappa$ with an antisymmetric component. In most such cases the complex vacuum, once it
appears at $c>c_{\rm lo}$, persists beyond the physical range of Eq.~(\ref{eq:ISS-c-bound}). In a
few of the samples, instead, the global maximum of $|\Tr(\kappa^T\Pi m)|$ is attained at
a real projector, and the vacuum is then real except within a finite window: at small $c$
the tree term selects the aligned real projector,
while at intermediate $c$ a complex projector with a slightly smaller
$|\Tr(\kappa^T\Pi m)|$ but a lower
tree-level cost is the global minimum. The reality of the maximizing projector therefore
does not by itself exclude a complex vacuum.

\begin{table}[htbp]
	\centering
	\begin{tabular}{c|cccc}
		 & $r=1.2$ & $r=2$ & $r=10$ & $r=100$ \\\hline
		complex vacuum & $13/50$ & $8/50$ & $3/50$ & $1/50$ \\
		$q_{25}(c_{\rm lo})$ & $0.047$ & $0.12$ & $0.19$ & $0.23$ \\
		${\rm median}(c_{\rm lo})$ & $0.088$ & $0.17$ & $0.22$ & $0.23$ \\
		$q_{75}(c_{\rm lo})$ & $0.20$ & $0.41$ & $0.32$ & $0.23$ \\
	\end{tabular}
	\caption{Numerical scan of the orientation selection by the messenger Coleman-Weinberg
	term. For $(N_c,N_f)=(3,4)$, a geometric mass spectrum $m=\diag(1,r,r^2,r^3)$
	(normalized; $c_{\rm lo}$ is scale invariant), 50 random Gaussian real $\kappa$, and a
	scan over $10^{-5}\lesssim c\lesssim2$ covering the range allowed by
	Eq.~(\ref{eq:ISS-c-bound}): the fraction of $\kappa$ whose global minimum of
	Eq.~(\ref{eq:ISS-vPi}) is a complex projector, and the quartiles (25th, 50th, 75th
	percentiles) of the lower edge $c_{\rm lo}$ of the complex window, evaluated over the
	$\kappa$ that admit one, so the sample size is that of the first row; for the $\kappa$
	that remain real at all $c$ no $c_{\rm lo}$ exists. The exactly degenerate case $r=1$
	is excluded from the scan: there the tree term of Eq.~(\ref{eq:ISS-vPi}) is flat and
	$c_{\rm lo}=0$ for the $\kappa$ that admit a complex vacuum. Both the fraction of
	complex vacua and the median $c_{\rm lo}$ improve monotonically as the degeneracy is lifted.}
	\label{tab:ISS-clo}
\end{table}

For the non-degenerate case, $F^M$ is in addition untouched by the shift of
Eq.~(\ref{eq:ISS-C-tadpole}), since Eq.~(\ref{eq:ISS-FM}) holds on the $\Phi_0$ block
whatever value $\Phi_0$ takes. A complex $\mc{C}$ would therefore reach $\arg M_3$ only
through the messenger mass matrix, to which the shifted meson VEV
$\vevx{M}$ contributes alongside the bare
mass. Using the equal-mass displacement as an order-of-magnitude estimate, we obtain
\begin{equation}
	M_3\propto\frac{1}{M_{\rm mess}}
	\sum_{i\in\text{remaining}}\frac{\kappa_{ii}\,\alpha|\Lambda|^2m_i}{M_*}
	\left(1-c_Z\,\frac{16\pi^2}{h^3}\,\frac{\sqrt\alpha\,|\Lambda|}{M_*}\,
	\frac{\mc{C}}{M_P^2 M_{\rm mess}}\sum_{j\in\text{remaining}}\kappa_{jj}+\cdots\right) .
	\label{eq:ISS-M3-mess}
\end{equation}
This is the same structure as Eq.~(\ref{eq:IYIT-M3-mess}), with $\sqrt\alpha|\Lambda|\sum_{j\in\text{remaining}}\kappa_{jj}/M_*$
in place of the unsuppressed $\kappa$ of the renormalizable portal there. The second
term carries $\arg\mc{C}$ through the messenger mass. At fixed gluino mass it is
independent of the messenger scale at this order
(Eq.~(\ref{eq:messenger-phase-size})). Its contribution to $\bar\theta$, including
the Higgs response, is given by Eq.~(\ref{eq:shift-floor}).
\section{The 3-2 model}
\label{sec:32model}

The Affleck-Dine-Seiberg 3-2 model is an $SU(3)\times SU(2)$ theory with chiral matter
$Q=(3,2)$, $\tilde u=(\bar3,1)$, $\tilde d=(\bar3,1)$, $L=(1,2)$
\cite{Affleck:1984uz,Affleck:1984xz,Shadmi:1999jy}. Since the matter content is chiral,
gauge invariance forbids a mass term, making this one of the simplest examples of genuinely
chiral dynamical SUSY breaking. The gauge-anomaly-free global $U(1)\times U(1)_R$ symmetry
is
\begin{center}
\begin{tabular}{c|cccc}
 & $SU(3)$ & $SU(2)$ & $U(1)$ & $U(1)_R$ \\\hline
$Q$      & $3$      & $2$ & $1/3$  & $1$  \\
$\tilde u$ & $\bar3$  & $1$ & $-4/3$ & $-8$ \\
$\tilde d$ & $\bar3$  & $1$ & $2/3$  & $4$  \\
$L$      & $1$      & $2$ & $-1$   & $-3$
\end{tabular}
\end{center}

\paragraph{Moduli space and gauge-invariant operators}
With no tree-level superpotential, the classical moduli space is parametrized by
$Q_{if}=\tilde Q^{if}=\diag(\alpha,\beta,0)$, $L=(0,\sqrt{\alpha^2-\beta^2})$,
and for generic $\alpha,\beta$ the gauge group is completely Higgsed. An equivalent, more convenient description uses the gauge-invariant
operators
\begin{equation}
  X_1=Q\tilde dL,\qquad X_2=Q\tilde uL,\qquad Y=\det(Q\tilde Q)
  \label{eq:32-invariants}
\end{equation}
(with $\tilde Q\equiv(\tilde u,\tilde d)$). The most general renormalizable superpotential
preserving $U(1)\times U(1)_R$ is
\begin{equation}
  W_{\rm tree}=\lambda\,Q\tilde dL=\lambda X_1
  \label{eq:32-W}
\end{equation}
alone (the $R$-charge of $X_2$ is $1-8-3=-10\neq2$, which is forbidden). Equation
(\ref{eq:32-W}) lifts every classical flat direction.

The two gauge angles can be shifted independently by anomalous field
redefinitions that preserve $\lambda X_1$ and the messenger portal through $X_1$
used below. Reversing the sign of $\tilde u$ shifts the $SU(3)$ angle by $\pi$.
Reversing the signs of $\tilde u$, $\tilde d$ and $L$ together shifts the
$SU(2)$ angle by $\pi$ and the $SU(3)$ angle by an integer multiple of $2\pi$.
For these interactions we can set both angles to zero in the field basis
used in the following calculation.

\paragraph{Nonperturbative superpotential}
Taking the weak-coupling limit $\lambda^2\ll g_2^2\ll g_3^2$ ($\Lambda_3\gg\Lambda_2$), the
minimum lies close to the D-flat direction, and the nonperturbative effects of $SU(2)$ are
exponentially suppressed relative to those of $SU(3)$ and can be neglected. $SU(3)$ confines
with $N_f=2$ flavors, and as the boundary ADS case with an integer power
$N_c-N_f=1$,
\begin{equation}
  W_{\rm np}=\frac{2\Lambda_3^7}{\det(Q\tilde Q)}=\frac{2\Lambda_3^7}{Y}
  \label{eq:32-Wnp}
\end{equation}
is generated by a single $SU(3)$ instanton (the general formula
Eq.~(\ref{eq:WADS}) with $N_c-N_f=1$). It is a single-valued function of $Y$.
The reality of the coefficient $2\Lambda_3^7$ is established by
the explicit one-instanton computation of Eq.~(\ref{eq:ADS-instanton})
\cite{Cordes:1985um,Finnell:1995dr} --- the collective-coordinate integral of a single
$SU(3)$ instanton at a completely Higgsed point is real and positive, and it simply
multiplies $\Lambda_3^7$, which is real at $\theta=0$\footnote{%
Here $\Lambda_3$ is normalized so as to make the coefficient $2$, i.e.\
$\Lambda_3^7=\Lambda^7/2$ with the $\Lambda$ of Eq.~(\ref{eq:ADS-instanton}).}.

\paragraph{Mechanism of SUSY breaking}
In the weak-coupling limit the light degrees of freedom are exhausted by $X_1,X_2,Y$, and
the full superpotential is given exactly by
\begin{equation}
  W=\frac{2\Lambda_3^7}{Y}+\lambda X_1 .
  \label{eq:32-Weff}
\end{equation}
Since $X_1$ and $Y$ are independent variables, $\partial W/\partial X_1=\lambda\neq0$
identically, regardless of the value of $Y$, so there is no F-flat point anywhere on the
moduli space: SUSY is broken. This F-flatness condition,
$\partial W/\partial X_1=0$, is a statement about $W$ alone and does not depend on the
K\"ahler normalization (unlike the physical auxiliary field $F^{X_1}$ of
Eq.~(\ref{eq:32-Fterm}), which does). The balance between the runaway of $W_{\rm np}$ in the $Y$ direction and the
tree-level term only stops the runaway and fixes the VEV scale of $Y$; what breaks SUSY is
the independent $X_1$-direction condition $\partial W/\partial X_1=\lambda\neq0$. This
conclusion is independent of the coupling hierarchy: for a generic hierarchy, a fourth
invariant $Z\equiv Q^3L$ and the $N_f=2$ quantum constraint of $SU(2)$, imposed through a
Lagrange multiplier $\mc{A}$, give the exact superpotential
$W=2\Lambda_3^7/Y+\mc{A}(Z-\Lambda_2^4)+\lambda X_1$ \cite{Shadmi:1999jy},
but $\partial W/\partial
X_1=\lambda\neq0$ is unchanged.

\paragraph{K\"ahler potential}
In the weak-coupling limit the K\"ahler potential is also known explicitly
\cite{Bagger:1994hh,Shadmi:1999jy}:
\begin{equation}
  K=24\,\frac{A+Bx}{x^2},\qquad
  A=\tfrac12\bigl(X_1^\dagger X_1+X_2^\dagger X_2\bigr),\quad B=\tfrac13\sqrt{Y^\dagger Y},
  \label{eq:32-Kahler}
\end{equation}
\begin{equation}
  x\equiv4\sqrt{B}\,\cos\!\left[\tfrac13\operatorname{Arccos}\!\Bigl(\frac{A}{B^{3/2}}\Bigr)\right],
  \qquad\text{the positive real root of }x^3-12Bx-16A=0\ (A\le B^{3/2}).
\end{equation}
The true scalar potential $V=W_i(K^{-1})^{i\bar\jmath}W_{\bar\jmath}^*$ must be evaluated
under this $K$ in the three variables $(X_1,X_2,Y)$. The approximation of a canonical
K\"ahler potential for a single modulus cannot be used: with a single chiral superfield
$\phi$ and a canonical K\"ahler potential, any root of $\partial W/\partial\phi=0$ is
necessarily a SUSY vacuum, so at least two independent fields, or a nontrivial K\"ahler
potential, are required to produce an isolated stable SUSY-breaking vacuum.

\paragraph{Known quantitative results}
By dimensional analysis and explicit minimization,
\begin{equation}
  v\sim\lambda^{-1/7}\Lambda_3,\qquad
  V_{\min}\sim\lambda^{10/7}\Lambda_3^4 .
  \label{eq:32-Vmin}
\end{equation}
The minimization of \cite{Bagger:1994hh} is carried out on the four real parameters that
remain once the global $U(1)$ and $U(1)_R$ are divided out, and returns a single minimum,
\begin{equation}
  X_1=0.50\,\frac{\Lambda_3^3}{\lambda^{3/7}},\qquad X_2=0,\qquad
  Y=2.58\,\frac{\Lambda_3^4}{\lambda^{4/7}} ,
  \label{eq:32-min}
\end{equation}
so $X_2=0$ at the minimum (the global $U(1)$ is preserved) and the set of minima is a
single $U(1)_R$ orbit. The massless spectrum
consists of three species: the goldstino, the Goldstone boson of the spontaneously
broken $U(1)_R$ (the R-axion), and a massless fermion of $U(1)$ charge $-2$ saturating the
global $U(1)$ anomaly \cite{Bagger:1994hh}\footnote{%
$Q,\tilde u,\tilde d,L$ are the quark sector and
lepton doublet of one Standard Model generation, but the
charged-lepton singlet $\bar E$ is missing; the massless fermion of charge $-2$ remaining in
the vacuum reproduces the anomalies $[U(1)]^3=-8$ and $U(1)$-gravity $=-2$ that the missing
$\bar E$ leaves behind.}.

The charge assignment in the table above is anomaly-free with respect to $SU(3)$
and $SU(2)$ (which is why $W_{\rm tree}$ and $W_{\rm np}$ are compatible under the same
$U(1)_R$), and it survives in the quantum theory as a continuous global symmetry.
The spontaneous breaking of $U(1)_R$ in the SUSY-breaking vacuum therefore produces a
continuous Goldstone direction (the R-axion), and
within the framework of global SUSY there is no dynamical mechanism to fix this phase (equivalently,
the overall phase of $\vevx{W}$).

\paragraph{Mechanism fixing the phase of the R-axion}
Within global SUSY the R-axion $a$ is exactly flat. Realizing the light fields nonlinearly
as $X_k=\vevx{X_k}\exp(ir_ka/f_R)$, with $f_R\sim v$ the R-axion decay constant, since $W$
carries overall $U(1)_R$ charge 2, one has
$W_{\rm dyn}(a)=\vevx{W_{\rm dyn}(0)}e^{2ia/f_R}$, dependent on $a$ in this way, while the
global-SUSY scalar potential itself, being $U(1)_R$-invariant, is independent of $a$.

Once coupled to supergravity, a constant must be added to the superpotential to cancel
the positive vacuum energy $|F|^2$ of the SUSY-breaking sector; its reality is what Sec.~\ref{sec:gaugino} establishes for a separate
pure $SU(2)$ sector at its vanishing hidden angle. In the setup of this paper, this is the
condensate contribution $\mc{C}$ defined in Eq.~(\ref{eq:C-source}). In the
condensate-dominated approximation, vanishing total vacuum energy
(Eq.~(\ref{eq:CC-condition})) fixes its magnitude through $3|\mc{C}|^2/M_P^2\simeq V_{\min}$, where
$V_{\min}$ of Eq.~(\ref{eq:32-Vmin}) is the global-SUSY vacuum energy of the 3-2 sector (its
$|F|^2$), and the gravitino mass is then
\begin{equation}
  |m_{3/2}|\simeq\frac{|\mc{C}|}{M_P^2}\simeq\sqrt{\frac{V_{\min}}{3}}\,\frac{1}{M_P}
  \label{eq:32-C}
\end{equation}
(the same relation used in Sec.~\ref{sec:IYIT-pseudo}).

$\mc{C}$ involves none of the fields of the 3-2 model ($X_1,X_2,Y$, and hence $a$) ---
gaugino condensation and the 3-2 model are decoupled sectors sharing no fields, and in
global SUSY there is no contact interaction between them. Hence
$\partial\mc{C}/\partial a\equiv0$ identically. Its value is fixed elsewhere, in
the gaugino-condensation sector of Sec.~\ref{sec:gaugino}, by the anomalous breaking
$U(1)_R\to Z_{2N_c}$ together with the spontaneous breaking
$\vevx{\lambda\lambda}\neq0$ there. As a charge-0 term added to the charge-2
superpotential $W_{\rm dyn}(a)$, it breaks the $U(1)_R$ of the 3-2 sector explicitly. This
breaking does not contradict the Nelson-Seiberg theorem \cite{Nelson:1993nf,Intriligator:2007cp} (proved
within global SUSY) --- the constant $\mc{C}$ makes no contribution at all to the
global-SUSY $F$-term equations that the theorem controls. This breaking becomes physically
relevant only once one promotes to the supergravity $F$-term condition
$D_iW=W_i+(K_iW)/M_P^2$.

The SUGRA scalar potential takes the standard form
\[
  V=e^{K/M_P^2}\bigl[K^{i\bar\jmath}D_iW\overline{D_jW}-3|W|^2/M_P^2\bigr].
\]
We assume that the supergravity contribution to the R-axion potential dominates
over the messenger-loop contribution.
For the angular potential we use $W\simeq W_{\rm dyn}(a)+\mc{C}$. In $|W|^2$,
the dependence on $a$ comes from the cross term
$2\,\mathrm{Re}[W_{\rm dyn}(a)^*\mc{C}]$. With
$W_{\rm dyn}(a)=\vevx{W_{\rm dyn}(0)}e^{2ia/f_R}$ and real coefficients
set by $\lambda,\Lambda_3^7,M_P$, the potential takes the form
\begin{equation}
  V_1(a)\;\approx\;-V_a\cos\!\Bigl(\frac{2a}{f_R}-\arg\mc{C}\Bigr),\qquad
  V_a\in\mathbb{R},\quad V_a\propto|\mc{C}| ,
  \label{eq:32-axion-pot}
\end{equation}
a genuine cosine potential \cite{Bagger:1994hh}. At tree level it is the only harmonic:
$V$ is sesquilinear in $W$ (each $D_iW$ is linear in $W$), so with
$W=W_{\rm dyn}(a)+\mc{C}$ the $a$-dependence enters only through
$\mc{C}\,\overline{W_{\rm dyn}(a)}$ and its conjugate. A term $e^{2iqa/f_R}$ with $q\ge2$
requires $q$ insertions of $\mc{C}$, since $W_{\rm dyn}$ carries $R$-charge 2 and $\mc{C}$
carries none; it arises only at higher order, from integrating out the radial modes
displaced by the $O(\mc{C})$ term or from loops, and is suppressed relative to
$\cos(2a/f_R)$ by $m_{3/2}$ over the mass $\sim\lambda v$ of those modes. This suppression
is a necessary input for what follows: a second harmonic larger than a quarter of the first
and of opposite sign would move the minimum to a CP-conjugate pair of points away from the
extrema below. With it, the potential has only two extrema,
$2a/f_R=\arg\mc{C}$ and $\arg\mc{C}+\pi$ (mod the period), at both of which
$\arg\vevx{W_{\rm dyn}}=\arg\mc{C}$ (mod $\pi$). For real $\mc{C}$ they are $a=0$ and
$a=\pi f_R/2$, and at both of them $\vevx{W_{\rm dyn}}$ is real (either
$+|\vevx{W_{\rm dyn}(0)}|$ or $-|\vevx{W_{\rm dyn}(0)}|$). The sign of $V_a$ (which extremum is
the true minimum) is model dependent, but regardless of the sign, both candidates are
real, so this reduces not to ``which of $N_c$-fold phases is real'' but to the qualitatively
safer binary question of ``which of the only two real candidates is the minimum.''

\paragraph{Reality of the $F$-term}
The reality of the $F$-term characterizing SUSY breaking in this model can also be checked
directly. What can be established without reference to the (non-canonical) K\"ahler metric
is the reality of the holomorphic derivatives $W_{X_1}\equiv\partial W/\partial X_1$
and $W_Y\equiv\partial W/\partial Y$, from which the physical auxiliary fields are built.
First, $W_{X_1}=\lambda$ is a constant,
independent of $X_1,Y,a$, and so is trivially real from the outset, before any R-axion
argument is invoked. Next, $W_Y$ depends on $a$. Since $R(X_1)=2$, $R(Y)=-2$, while
$\Lambda_3^7$ --- the quantity that Eq.~(\ref{eq:Lambda-b-phase}) defines, here with
$b=3N_c-N_f=7$, and which is therefore real at $\theta_3=0$ --- is neutral under the
anomaly-free $U(1)_R$ and so does not rotate under it, one has $Y(a)=Y(0)\,e^{-2ia/f_R}$ and hence
\begin{equation}
  W_Y(a)=-\frac{2\Lambda_3^7}{Y(a)^2}=W_Y(0)\,e^{4ia/f_R} ,
\end{equation}
and at the extrema $a=0,\ \pi f_R/2$ of Eq.~(\ref{eq:32-axion-pot}) for real $\mc{C}$, both give
$e^{4ia/f_R}=1$, so $W_Y$ is real at both extrema. For real $\lambda$ and
$\Lambda_3^7$ the global-SUSY potential satisfies $V(\bar\phi)=V(\phi)$, so complex
conjugation maps its set of minima to itself, and by Eq.~(\ref{eq:32-min}) that set is a
single $U(1)_R$ orbit. Conjugation acts on that circle as a reflection with exactly two
fixed points, $a=0$ and $a=\pi f_R/2$, and the potential of
Eq.~(\ref{eq:32-axion-pot}), CP-even for real $\mc{C}$, is extremized there. This reality holds within the exact-CP
assumption that runs through this entire paper (that every parameter in the UV Lagrangian is
taken real).

The physical auxiliary fields follow from these by
$F^i=-K^{i\bar\jmath}\overline{W_{\jmath}}$, with $K_{i\bar\jmath}$ the (non-canonical)
K\"ahler metric of Eq.~(\ref{eq:32-Kahler}) ($X_2=0$ at the vacuum, so only the $X_1,Y$ block
matters). $K$ itself is built from the $U(1)_R$-invariant bilinears $X_1^\dagger
X_1,X_2^\dagger X_2,Y^\dagger Y$ and does not depend on the R-axion phase $a$, while the
metric carries the $R$-charges of its indices: the diagonal entries are $a$-independent, and
the off-diagonal entry $K_{X_1\bar Y}\propto\bar X_1Y$ rotates as $e^{-4ia/f_R}$. The
$a$-dependence of each auxiliary field is fixed by its $R$-charge, $R(F^i)=R(\phi^i)-2$, which
is $0$ for $X_1$ and $-4$ for $Y$:
\begin{equation}
  F^{X_1}(a)=F^{X_1}(0),\qquad F^Y(a)=F^Y(0)\,e^{-4ia/f_R} .
  \label{eq:32-FX1-FY}
\end{equation}
In $F^{X_1}$ the diagonal term involves the constant $W_{X_1}=\lambda$, and in the
off-diagonal term the phase of $K^{X_1\bar Y}$ cancels against that of $\overline{W_Y}$. At
$a=0$, the fixed point of complex conjugation identified above, $X_1$, $Y$, $K_{i\bar\jmath}$
and $W_i$ are all real, so $F^{X_1}(0)$ and $F^Y(0)$ are real. Hence $F^{X_1}$ is real at
every point of the R-axion orbit, independently of where the R-axion sits, and $F^Y$ is real
at the two extrema $a=0,\ \pi f_R/2$ of Eq.~(\ref{eq:32-axion-pot}), where
$e^{-4ia/f_R}=1$.

\paragraph{Coupling to messengers}
Since none of $Q,\tilde u,\tilde d,L$ is a gauge singlet, a messenger portal must couple
through the gauge-invariant combination $X_1=Q\tilde dL$ rather than through the elementary
fields directly, and since $X_1$ has mass dimension 3 the renormalizable portal available in
Sec.~\ref{sec:IYIT-pseudo} is not available here; the leading portal is the composite-operator
type of \cite{Murayama:2007fe}, now for $X_1$ rather than a meson,
$W\supset M_{\rm mess}f\bar f+X_1f\bar f/M_*^2$. The induced gaugino mass is
\begin{equation}
  M_3\propto\frac{F^{X_1}/M_*^2}{M_{\rm mess}+\vevx{X_1}/M_*^2} ,
  \qquad
  F^{X_1}=-K^{X_1\bar X_1}\overline{W_{X_1}}-K^{X_1\bar Y}\overline{W_Y} ,
  \label{eq:32-Fterm}
\end{equation}
where the two terms on the right are parametrically comparable, both
$\sim\lambda^{3/7}\Lambda_3^4$ for $v\sim\lambda^{-1/7}\Lambda_3$\footnote{%
With $X_2=0$ at the vacuum, $\vevx{X_1}\sim v^3$ and $\vevx{Y}\sim v^4$, so
$A=\tfrac12|X_1|^2\sim v^6$ and $B^{3/2}\sim(v^4)^{3/2}=v^6$: the ratio $A/B^{3/2}$ entering
Eq.~(\ref{eq:32-Kahler}) is $O(1)$ independently of $v$, the $\operatorname{Arccos}$ carries
no small parameter, and $x\sim\sqrt B\sim v^2$, $K\sim v^2$. Hence
$K_{X_1\bar X_1}\sim v^{-4}$ and $K_{X_1\bar Y}\sim v^{-5}$, so the inverse metric has
$K^{X_1\bar X_1}\sim v^4$ and $K^{X_1\bar Y}\sim v^5$. With $W_{X_1}=\lambda$ and
$W_Y\sim\Lambda_3^7/v^8$ this gives $K^{X_1\bar X_1}\overline{W_{X_1}}\sim\lambda v^4$ and
$K^{X_1\bar Y}\overline{W_Y}\sim\Lambda_3^7/v^3$, both equal to $\lambda^{3/7}\Lambda_3^4$
at $v=\lambda^{-1/7}\Lambda_3$. The $A$-$B$ mixing in $K$ is therefore unsuppressed.}:
the off-diagonal
$K^{X_1\bar Y}\overline{W_Y}$ piece cannot be dropped from the magnitude of $F^{X_1}$, and
its phase is settled by Eq.~(\ref{eq:32-FX1-FY}): $F^{X_1}$ is real on the whole R-axion
orbit. $M_*,M_{\rm mess}$ are real by the CP invariance of Eq.~(\ref{eq:setup}), and
$\vevx{X_1}\sim\lambda^{-3/7}\Lambda_3^3$ (from $X_1=Q\tilde dL\sim v^3$) rotates as
$e^{2ia/f_R}$, so it takes the real values $\pm|\vevx{X_1}|$ precisely at the two extrema
fixed by the real $\mc{C}$; there $M_3$ is real. The reality of $M_3$ thus relies on fixing
the R-axion through the messenger mass alone. A portal through $Y$ would instead transmit
$F^Y$, which rotates with $a$ by Eq.~(\ref{eq:32-FX1-FY}); there the R-axion must be fixed
for the $F$-term itself as well, and $F^Y$ and $\vevx{Y}$ are both real at the same two
extrema. Under the field redefinition shifting the $SU(3)$ angle by $\pi$,
$Y$ changes sign, and the coefficient of a portal through $Y$ changes sign
in the new basis.

\section{Conclusion and discussion}
\label{sec:conclusion}

DSB and gaugino condensation are strongly-coupled confining dynamics, and even when every
UV coupling is real they can introduce a CP phase into the constant term and the
SUSY-breaking $F$-term. The phase arises when the real Lagrangian does not fix a real
vacuum: the vacuum is either one of a set of degenerate vacua related by a spontaneously
broken discrete symmetry, or a complex point selected by the potential itself (IYIT) or by
the loop corrections (ISS). For
pure SYM the condensate branches come from the $N_c$-th root of the
holomorphic scale. For IYIT the tree-level $F$-term potential
selects a purely imaginary vacuum for half of the real Yukawa couplings, the CP-conjugate
pair being related by the $Z_4$ center; for ISS the $F$-term potential leaves the
orientation of the vacuum degenerate, and it is the messenger loops that select it.
Whether such complex vacua appear is
a property of the vacuum equations, or, where they leave an orientation degenerate, of the
loop-corrected effective potential; a tree-level polynomial superpotential is no
exception. The analysis of IYIT, ISS and the 3-2 model examines the holomorphic
branch structure and the vacuum selected in each case.

\paragraph{Channel (A): the constant term}
When pure SYM supplies $\mc{C}$, the condensate is real on every branch
for $SU(2)$ at $\theta=0$. At $\theta=\pi$ the same theory has two purely
imaginary condensates. A complex $\mc{C}$ reaches the gluino mass
through anomaly mediation, with the benchmark constraint in
Eq.~(\ref{eq:m32-F-bound}). For a bare real $\mu$, the compensator phase
also enters the quark determinant through the Higgs vacuum expectation
values at tree level. Equation~(\ref{eq:Higgs-phase}) can then give the
stronger constraint.
In massless SQCD, an independent soft term with a real coefficient can fix
the phase of the ADS condensate at zero or $\pi$ (Sec.~\ref{sec:ADS}).

\paragraph{Channel (B): the DSB constructions}
The phase of the $F$-term reaches $\arg M_3$ without dilution; its contribution to
$\bar\theta$ cancels only if the phases of $M_3$ and $b$ are fully aligned, including
compensator contributions. Otherwise, it has to be assessed model by model. The three
constructions studied here can be safe, each for its own reason.
\begin{itemize}
\item IYIT (Sec.~\ref{sec:IYIT}) rests on the $N_f=N_c$ quantum moduli constraint, in
  which $\Lambda^{2N_c}$ enters linearly with a real coefficient and no root taken
  (Secs.~\ref{sec:ARV}--\ref{sec:cancel}). At $m=0$, where IYIT sits, the meson carries no
  discrete branch. However, the point of the constraint surface selected as the vacuum need
  not be real: depending on the real Yukawa coupling, the meson VEV is either real or purely
  imaginary, and in the latter case CP is spontaneously broken. In the real class the real
  $\mc{C}$ displaces the pseudomodulus from the origin along a real direction. The $SU(2)$
  $\theta$ angle enters this sector only through the phase of $\Lambda^4$ in
  $\Pf M=\Lambda^4$: $\theta=0$ and $\theta=\pi$ both give a real $\Lambda^4$, of opposite
  sign, and exchange the real and purely imaginary classes at fixed real Yukawa couplings.
\item ISS (Sec.~\ref{sec:ISS}) is governed by a tree-level polynomial superpotential
  in the magnetic dual. Computed in electric variables, the SUSY-breaking order parameter is
  real whenever the quark-mass splittings exceed the messenger-loop corrections: the
  convention-dependent coefficient of the
  dual superpotential cancels, and the holomorphic scale enters only through the real
  K\"ahler metric. The one term that carries a branch vanishes at the metastable vacuum, where
  the magnetic gauge group is completely Higgsed, and the supergravity displacement of $\Phi$
  leaves it Higgsed. When the splittings fall below the messenger-loop corrections the
  orientation of the vacuum is instead fixed by the messenger loops, and a portal coupling with an
  antisymmetric part can select a complex orientation and generate an $\mathcal{O}(1)$ phase;
  this is the exception noted in the abstract.
\item In the 3-2 model (Sec.~\ref{sec:32model}), a real external constant fixes the
  R-axion at a CP-conserving point.
\end{itemize}

\paragraph{A second route for the phase of $\mc{C}$}
The constant term displaces the pseudomodulus in IYIT and ISS and fixes the R-axion in
the 3-2 model. Through these effects, a complex $\mc{C}$ can induce a phase in the
messenger mass and, for the portal through $Y$ in the 3-2 model, in the $F$-term itself.
In IYIT and ISS the messenger mass phase persists at fixed gluino mass as the messenger
scale is lowered. Its contribution
to $\bar\theta$ also depends on the Higgs bilinear response
(Eq.~(\ref{eq:shift-floor})). In the real-vacuum cases discussed above, a real constant
term keeps this contribution real, provided the imaginary part of the remaining vacuum
superpotential is sufficiently small, as assumed in Sec.~\ref{sec:M3channels}.

\section*{Acknowledgments}
N.Y. is supported by a start-up grant from Zhejiang University.
We acknowledge the use of Claude Code (Anthropic), with Claude Opus 5-series and Claude Fable 5.1 models, for literature and reference checking, for assistance in verifying the analytic derivations in Secs.~\ref{sec:M3channels}--\ref{sec:32model}, and for editorial assistance in preparing the manuscript, and of DeepSeek-V4.1-Flash (DeepSeek) and Gemini 3.1 Pro (Google) for critical review of the manuscript.
We also used OpenAI's Astra and Sol models through Codex for editorial assistance with the manuscript.
All derivations and conclusions were directed and independently verified by the author, who takes full responsibility for the content of this paper.

\appendix
\numberwithin{equation}{section}

\section{Derivation of the selection rule}
\label{app:selection}

This appendix derives Eq.~(\ref{eq:selection}). Let $n_\lambda$, $n_\varphi$ and $n_\psi$ be
the numbers of gauge-invariant bilinears $\lambda\lambda$, $\tilde\varphi\varphi$ and
$\tilde\psi\psi$ in $G$, and $n_F$ the number of field-strength factors, in the $\nu$-instanton sector. Three conditions fix these
numbers \cite{Amati:1984uz}.

\textit{(i) Anomalous $U(1)_R$.} $U(1)_R$ is the $R$-symmetry with $R(Q)=R(\tilde Q)=0$: $\lambda\to e^{i\alpha}\lambda$,
$\psi,\tilde\psi\to e^{-i\alpha}\psi,\tilde\psi$, with the squarks $\varphi,\tilde\varphi$
neutral. The $\nu$-instanton carries $2N_c\nu$ gluino zero modes and $N_f\nu$ zero modes
each of $\psi$ and $\tilde\psi$, so it violates this charge by
$\Delta R=2N_c\nu-2N_f\nu$. Since $\lambda\lambda$, $\tilde\varphi\varphi$ and
$\tilde\psi\psi$ carry charges $+2$, $0$ and $-2$, $G$ is nonzero only if
\begin{equation}
  n_\lambda-n_\psi=(N_c-N_f)\,\nu .
  \label{eq:sel-R}
\end{equation}

\textit{(ii) Anomalous $U(1)_A$.} $U(1)_A$ is the common phase rotation of $Q$ and $\tilde Q$,
under which $\varphi,\tilde\varphi,\psi,\tilde\psi$ carry charge $+1$ and $\lambda$ is
neutral. The same zero modes violate it by $\Delta A=2N_f\nu$, and since
$\tilde\varphi\varphi$ and $\tilde\psi\psi$ carry charge $+2$ while $\lambda\lambda$ is
neutral,
\begin{equation}
  n_\varphi+n_\psi=N_f\,\nu .
  \label{eq:sel-A}
\end{equation}

\textit{(iii) Mass dimension.} In the massless theory the constant $G$ can obtain its mass
dimension only from the renormalization-group invariant $\Lambda^{b\nu}$ with $b=3N_c-N_f$,
which is how the instanton weight $e^{-8\pi^2\nu/g^2}$ combines with the scale of the
't~Hooft determinant. Hence
\begin{equation}
  3n_\lambda+2n_\varphi+3n_\psi+2n_F=(3N_c-N_f)\,\nu .
  \label{eq:sel-dim}
\end{equation}

Substituting Eqs.~(\ref{eq:sel-R}) and (\ref{eq:sel-A}) into Eq.~(\ref{eq:sel-dim}) gives
$4n_\psi+2n_F=0$, so $n_\psi=n_F=0$, and Eqs.~(\ref{eq:sel-R}) and (\ref{eq:sel-A}) reduce to
Eq.~(\ref{eq:selection}).

\section{The quantum constraint from $SU(4)_F$ covariance for $N_c=2$}
\label{app:SU4}

This appendix gives the details behind the averages over the vacua $\vevx{\det M}=2\gamma\Lambda^4$ and $\vevx{B\tilde B}=-\gamma\Lambda^4$ used in Sec.~\ref{sec:ARV}.

\paragraph{Vacuum average of $\det M$}
In the clustering limit the $N_c$-point meson Green's function of Sec.~\ref{sec:ARV}
becomes
\begin{equation*}
  \sum_\Omega w_\Omega\prod_k\vevx{\Omega|M^{i_k}_{j_k}|\Omega}
  =\epsilon^{i_1\cdots i_{N_c}}\epsilon_{j_1\cdots j_{N_c}}\gamma\Lambda^{2N_c} .
\end{equation*}
Contracting with $\epsilon_{i_1\cdots i_{N_c}}\epsilon^{j_1\cdots j_{N_c}}/N_c!$ gives
\begin{equation}
  \vevx{\det M}\equiv\sum_\Omega w_\Omega\det M(\Omega)
  =\frac{(N_c!)^2}{N_c!}\,\gamma\,\Lambda^{2N_c}=N_c!\,\gamma\,\Lambda^{2N_c} .
\end{equation}

\paragraph{Identification of $M^i_j$, $B$, $\tilde B$ with $\hat M^{ab}$}
For $N_c=2$, $\tilde Q$ is also a doublet, and $Q_1,Q_2,\tilde Q^1,\tilde Q^2$ form four
doublets $Q^a=(Q_1,-Q_2,\tilde Q^1,\tilde Q^2)$, with the sign of $Q_2$ chosen so that
$\det M-B\tilde B=+\Pf\hat M$. Taking $a=1,2$ for $Q_i$ and $a=3,4$ for $\tilde Q$, we identify
\begin{equation}
  M^1_1=\hat M^{13},\quad M^2_2=-\hat M^{24},\quad M^1_2=\hat M^{14},\quad
  M^2_1=-\hat M^{23},\quad B\tilde B=-\hat M^{12}\hat M^{34},
\end{equation}
so that $\det M=-\hat M^{13}\hat M^{24}+\hat M^{14}\hat M^{23}$ and
\begin{equation}
  \det M-B\tilde B=\hat M^{12}\hat M^{34}-\hat M^{13}\hat M^{24}+\hat M^{14}\hat M^{23}
  =\Pf\hat M .
\end{equation}

\paragraph{The two-point function}
$\hat M^{ab}(x)$ is the lowest component of a chiral superfield, so its two-point function is
constant, and the one-instanton measure is invariant under the anomaly-free $SU(4)_F$. The
unique invariant tensor with two antisymmetric pairs of $SU(4)_F$ indices is
$\epsilon_{abcd}$, so $\vevx{0|\hat M^{ab}(x_1)\hat M^{cd}(x_2)|0}=\kappa\,\epsilon_{abcd}\Lambda^4$.
With the identification above,
$\vevx{M^1_1M^2_2}=-\kappa\epsilon_{1324}\Lambda^4=\kappa\Lambda^4$ and
$\vevx{M^1_2M^2_1}=-\kappa\epsilon_{1423}\Lambda^4=-\kappa\Lambda^4$, which agree with
$\epsilon^{i_1i_2}\epsilon_{j_1j_2}\gamma\Lambda^4$ for $\kappa=\gamma$.

\paragraph{Averages}
Each of the three terms of $\Pf\hat M$ averages to the same value,
\begin{equation}
  \vevx{\hat M^{12}\hat M^{34}}=\gamma\epsilon_{1234}\Lambda^4,\quad
  -\vevx{\hat M^{13}\hat M^{24}}=-\gamma\epsilon_{1324}\Lambda^4,\quad
  \vevx{\hat M^{14}\hat M^{23}}=\gamma\epsilon_{1423}\Lambda^4,
\end{equation}
all equal to $\gamma\Lambda^4$. $\det M$ consists of two of these terms and $-B\tilde B$ of the
third, so
\begin{equation}
  \vevx{\det M}=2\gamma\Lambda^4,\qquad \vevx{B\tilde B}=-\gamma\Lambda^4,\qquad
  \vevx{\det M-B\tilde B}=\vevx{\Pf\hat M}=3\gamma\Lambda^4 ,
\end{equation}
where $\vevx{\det M}$ agrees with the general-$N_c$ value $N_c!\,\gamma\Lambda^{2N_c}$ above.

\bibliographystyle{utphys}
\bibliography{refs}

\end{document}